\documentclass[peerreview,nodraftcls]{IEEEtran}
\usepackage{graphicx,subfig,amsthm,amsmath,latexsym,amssymb}
\usepackage{float,epsfig,multirow,rotating,verbatim,wrapfig,cite,booktabs}
\usepackage{color,xr,array,hyperref}

\ifCLASSINFOpdf
\else
\fi
\def\bSig\mathbf{\Sigma}

\newcommand{\bd}{\boldsymbol{d}}
\newcommand{\blambda}{\boldsymbol{\lambda}}
\newcommand{\balpha}{\boldsymbol{\alpha}}

\newcommand{\bDelta}{\boldsymbol{\Delta}}

\newcommand{\bbeta}{\boldsymbol{\beta}}

\newcommand{\bGamma}{\boldsymbol{\Gamma}}
\newcommand{\bgamma}{\boldsymbol{\gamma}}
\newcommand{\bSigma}{\boldsymbol{\Sigma}}

\newcommand{\bpsi}{\boldsymbol{\psi}}

\newcommand{\bmu}{\boldsymbol{\mu}}

\newcommand{\bvarphi}{\boldsymbol{\varphi}}

\newcommand{\bOmega}{\boldsymbol{\Omega}}

\newcommand{\bA}{\boldsymbol{A}}
\newcommand{\bB}{\boldsymbol{B}}
\newcommand{\bC}{\boldsymbol{C}}
\newcommand{\bc}{\boldsymbol{c}}
\newcommand{\bD}{\boldsymbol{D}}
\newcommand{\bff}{\boldsymbol{f}}
\newcommand{\bV}{\boldsymbol{V}}

\newcommand{\bM}{\boldsymbol{M}}

\newcommand{\bI}{\boldsymbol{I}}

\newcommand{\bJ}{\boldsymbol{J}}
\newcommand{\bK}{\boldsymbol{K}}
\newcommand{\bk}{\boldsymbol{k}}
\newcommand{\bS}{\boldsymbol{S}}

\newcommand{\bQ}{\boldsymbol{Q}}

\newcommand{\bW}{\boldsymbol{W}}

\newcommand{\bU}{\boldsymbol{U}}
\newcommand{\bx}{\boldsymbol{x}}

\newcommand{\by}{\boldsymbol{y}}

\newcommand{\bz}{\boldsymbol{z}}

\newcommand{\bzero}{\boldsymbol{0}}

\newcommand{\bone}{\boldsymbol{1}}

\newcommand{\mbC}{\boldsymbol{\mathcal C}}

\newcommand{\mR}{\mathcal R}

\newcommand{\argmin}{\operatornamewithlimits{argmin}}
\newcommand{\E}{\mathbb{E}}

\begin{document}
%
\title{Efficient Bandwidth Estimation in Two-dimensional Filtered Backprojection Reconstruction} 
%
%
%
\author{Ranjan~Maitra
\thanks{R. Maitra is Professor in the Department of
  Statistics at Iowa State University,  Ames, Iowa, USA. 
  This  research was supported, in part, by the National Science
  Foundation (NSF) under its CAREER Grant No. DMS-0437555 and the National 
  Institute of Biomedical Imaging and Bioengineering~(NIBIB) of the
  National Institutes of Health (NIH) under its Award No. R21EB016212. The
  content of this paper is solely the responsibility of the author and
  does not represent the official views of the NSF, NIBIB or NIH.}
}

\newtheorem{theorem}{Theorem}
\newtheorem{lemma}[theorem]{Lemma}
\newtheorem{corollary}[theorem]{Corollary}
\newtheorem{definition}[theorem]{Definition}

\maketitle
\thispagestyle{empty}
\setcounter{page}{1}

\newcommand{\I}{{\textrm{I}}}
\newcommand{\avg}[1]{{<\!\!#1\!\!>}}
\newcommand{\citep}{\cite}
\newcommand{\citet}{\cite}

\begin{abstract}
  A generalized cross-validation approach to estimate the
  reconstruction filter bandwidth in two-dimensional Filtered Backprojection is 
  presented. The method writes the reconstruction equation in
  equivalent backprojected  filtering form, derives results on
  eigendecomposition of symmetric two-dimensional circulant matrices
  and applies them to make bandwidth estimation a computationally
  efficient operation within the context of standard
  backprojected filtering reconstruction. Performance evaluations on a
  wide range of  
simulated emission tomography experiments give promising results. The
superior performance holds at both low and high total expected counts,
pointing to the method's applicability even in weaker
signal-noise situations. The approach also applies to the more general
class of elliptically symmetric   filters, with reconstruction
performance often better than even that obtained with the true optimal
radially symmetric filter. 
\end{abstract}

\section{Introduction}
\IEEEPARstart{F}{iltered} Backprojection (FBP)~\citep{bracewellandriddle67,ramachandranandlakshminarayanan71,kakandslaney88} is commonly used in
tomographic reconstruction where the goal is to estimate an object or
emitting source distribution from its degraded linear
projections that have been recorded by an appropriately designed
set of
detectors~\citep{mersereauandoppenheim74,bruyant02,herman09}. 
Such scenarios arise in areas such as  astronomy
~\citep{naandlee94,starckandmurtagh06,haefneretal15}, materials science and
non-destructive 
evaluation~\citep{nagataetal95,wrightetal97,ladeetal05,zhaoetal11,myagotinetal13},
electron microscopy~\citep{prabhatetal17} and tomosynthesis~\citep{sechopoulos13} or in object detection
with security scanners~\citep{megherbietal13}. A popular
application is in emission tomography imaging such as Single Photon
Emission Computed Tomography (SPECT) or Positron Emission Tomography
(PET)~\citep{terporgossianetal80,haweetal12,osullivan06,wolsztynskietal17}
that forms the primary setting for this article. The
challenges in emission tomography inherent in the dosimetry constraints and
the Poisson distribution of the sinogram emissions have meant the
development of sophisticated statistical   
methods~\citep{vardietal85,green90,hudsonandlarkin94,nuytsandfessler03,staymanandfessler04,osullivanandosuilleabhaein13}. Nevertheless,
the computationally fast FBP is still very commonly used. Further, many of the  
gains associated with some of the sophisticated methods are typically
in background regions and easily recovered by a quick
postprocessing of the reconstructions~\citep{osullivanetal93,osullivan95}.  Also, three-dimensional (3D) PET reconstructions are often obtained from 2D 
sinograms acquired with septa in place or with  Fourier Rebinning
(FORE)~\citep{defriseetal97,daube-witherspoonetal03,leeetal04}. However,
FBP reconstruction is generally accompanied by smoothing that involves a
bandwidth or resolution size parameter, often specified in 
terms of its full-width-at-half-maximum (FWHM), that must ideally be
optimally set to get spatially consistent reconstructions. Similar to
nonparametric function estimation in statistics, the quality of
reconstruction is evaluated by, for instance, the squared error 
loss function~\citep{hall92,nychkaandcox89,silverman86,stone84,wahba90}.

Data-dependent unbiased risk estimation
techniques~\citep{girard87,pawitanandosullivan93} -- with 
 practical modifications~\citep{osullivanandpawitan96} to adjust for
 the extra-Poisson variation in  corrected PET data -- have been
 developed. The methodology is interpretable as a form of
 cross-validation (CV). 
Many practitioners however forego bandwidth selection schemes that
involve additional steps beyond  reconstruction, and instead either
use a value that is fixed or  visually chosen 
and typically undersmooths reconstructions. 

The use of CV~\citep{geisser75,stone74} and the rotationally
invariant Generalized CV (GCV)~\citep{golubetal79} is quite prevalent
for bandwidth selection in nonparametric function estimation~\citep{cravenandwahba79} and  image 
restoration~\citep{reevesandmerserau90,reeves92,nguyenetal01}.
For moderate sample sizes, CV-obtained bandwidth parameters yield 
the best smoothed 
linear ridge and nonparametric
regression estimators~\citep{hallandtitterington87}.  
In image deblurring where a degraded version of the true image after
convolving  via a point-spread function is observed,  
\citet{thompsonetal91} and \citet{reeves92} provide optimal GCV
bandwidths that usually perform 
well~\citep{thompsonetal91} -- howbeit see ~\citet{thompsonetal89} for
examples of undersmoothing -- but  are impractical to obtain in
tomographic applications because they require indirect function
estimation (see Section 4.1 of \citep{reevesandmerserau90}).
Consequently, Section~\ref{methodology} of this article shows that the
singular value decomposition~(SVD) of the 
reconstruction (in matrix notation) can  be readily
obtained from results on symmetric one-dimensional (1D) and 2D
circulant matrices, which are also derived here. The Predicted Residual Sums of
Squares~(PRESS) are then very easily obtained in a
similar spirit to \citet{golubetal79} and practically minimized en
route to FBP reconstruction to 
obtain the GCV-estimated bandwidth. The methodology is evaluated 
on simulated 2D phantom data in
Section~\ref{experiments}. Our implementation and results show that
GCV selection and PET reconstruction can be carried out in less than a
second, achieving an integrated squared error that is very close to the
ideal. Moreover, our optimal reconstructions have the maximum relative
benefits at lower rates of emissions.
 Further, the methodology can be used to optimally select parameters
 in the wider class of elliptically symmetric 2D kernel smoothers.  
Postprocessing proposed in \citet{osullivanetal93} further improves
reconstruction quality by removing negative artifacts. 
Our article concludes with some discussion including areas that could
benefit from further extensions of our development.   
This paper also has a supplement.

\section{Theory and Methods}
\label{methodology}
\subsection{Background and Preliminaries}
Let $ y_{r\theta}$ be the attenutation-, scatter- and
randoms-corrected sinogram measurement along the line of response (LOR)
indexed by $(r,\theta)$,
$r=1,2,\ldots,R;\theta=1,2\ldots,\Theta$. Assume that the sinogram has
$n=R\Theta( > p)$ LORs. Suppose that we use FBP to reconstruct the
underlying source distribution in an imaging grid of $p$ pixels.  
In convolution form, the $i$th FBP-reconstructed pixel value, for
 $i = 1, 2, \ldots, p$, is
\begin{equation}
\label{reconeqn}
\hat{\lambda}_i^h =
\sum_{\theta=1}^\Theta\sum_{r=1}^R e_h(x_i\cos\theta+y_i\sin\theta -
r)y_{r\theta}.
\end{equation}
Here,  $e_h(\cdot)$ is the 
convolution filter with FWHM $h$. The summation over $r$ is a
convolution and efficiently achieved though a series of 1D discrete
Fast Fourier Transforms 
(FFT) and linear interpolation while the summation over $\theta$ is
the slower backprojection step. The Projection Slice Theorem and
properties of the Radon transform show that there is an equivalent
form of FBP called Backprojected Filtering (BPF)  where the
backprojection step is applied first and is followed by 2D convolution
in the imaging
domain~\citep{merserauandoppenheim74,natterer86,deans93}. BPF
reconstructions have an equivalent 
characterization~\citep{osullivanetal93} as a smoothed least-squares (LS)
solution in matrix form as
\begin{equation}
\hat\blambda^{h} = \bS_{ h }(\bK'\bK)^{-1}\bK'\by,
\label{matrecon}
\end{equation}
where $\by$ is the $n$-dimensional vector of corrected Poisson data in
the sinogram domain, $\bK$ is a discretized version of the Radon transform and
$\bS_{ h }$ is a smoothing matrix with FWHM $ h $. The application
of $\bK'$ to $\by$ is {\em backprojection} and the multiplication by
$(\bK'\bK)^{-1}$ is filtering and can be done using FFTs because the
matrix $(\bK'\bK)^{-1}$ is approximately 2D circulant. Moreover, if
$\bS_{ h }$ is also  2D circulant, the operation $\bS_{ h }
(\bK'\bK)^{-1}$ can be done in one convolution step. 

\subsubsection*{Comments} We make a few remarks on our setup:
\paragraph{Smoothed FBP}
A reviewer has  pointed out that original FBP does not
incorporate any smoothing and that our development here really 
pertains to smoothed FBP reconstructions. We agree but drop the
qualifier in smoothed FBP for brevity and also because it is hard to 
conceive using FBP without smoothing in a practical setting because of
the lack of spatial consistency in unsmoothed FBP reconstructions. 
\paragraph{Choice of $\bS_h$}
Another reviewer has asked about the assumption of $\bS_h$ being a
circulant matrix. FBP/BPF mostly use radially symmetric smoothing
filters that are 2D circulant,
so we do not consider this restriction to be a major limitation. At
this point, we consider $\bS_h$ that arises
from a radially symmetric Gaussian kernel, with $(k,j)$th element $S_h(k,j)
\propto \exp{-(k^2+j^2)/2h^2}$. (Section~\ref{elliptical} further
widens our class of reconstruction filters to include 
elliptically symmetric kernels.) 

This paper develops an optimal method to estimate $h$ in the setup of \eqref{reconeqn}.
Leave-one-out CV (LOOCV) is often used to choose the optimal $h$ in density
estimation~\citep{silverman86,wandandjones95}. For FBP, a LOOCV
strategy would remove 
$y_j\equiv y_{r,\theta}$ for the $j$th LOR $(r,\theta)$, obtain an
estimate of $\hat\blambda_{-j}^h \equiv \hat\blambda^h_{-(r,\theta)}$
from the remaining LOR data ($y_{-j}$), project it along the LOR and
compare the projected (predicted) value with the (observed) $y_j$
in terms of its squared error. 
LOOCV leads to the PRESS statistic 
\begin{equation}
\mathcal P(h;\by)= \sum_{j=1}^n [(\bK\hat\blambda_{-j}^ h )_j-y_j]^2,
\label{PRESS}
\end{equation} 
where $(\bK\hat\blambda_{-j}^h)_j$ is the $j$th coordinate of
the expected emissions predicted from the leave-$j$th-LOR-out
reconstruction $\hat\blambda_{-j}^h$ (obtained from $y_{-j}$) and  is
$\bk_j'\hat\blambda_{-j}^h$ with  $\bk_j'$ denoting the $j$th row of
$\bK$. Minimizing \eqref{PRESS} over $ h $, that is, finding
 $\argmin_h\mathcal P(h;\by)$
involves multiple evaluations, for each
$h$, of\eqref{PRESS}, with each calculation requiring $n$ 
reconstructions and projections (one for each left-out LOR) without the
benefit of the FFT because removing a LOR  damages the circulant
structure of $\bK'\bK$, and 
choosing the $h$ minimizing   \eqref{PRESS}.
Such an approach, with time-consuming calculations for each $h$,  is
computationally impractical, so we derive an invariant version of
\eqref{PRESS} that reduces to an easily computed function of  $ h $.      

\subsection{An Invariant PRESS Statistic and GCV Estimation of $h$}
\label{press}
To obtain a GCV estimate of $h$, we first state and prove our
\begin{theorem}
\label{gcv.theorem}
Let $\bU=[\bU_1 \vdots \bU_2]$ be the $n\times n$ orthogonal matrix of 
the left singular vectors of $\bK$, with $\bU$ 
partitioned into matrices $\bU_1$ and $\bU_2$ with $p$ and $n-p$
columns, respectively. Also, let $\bOmega_h$ be the diagonal matrix of
the $p$ eigenvalues of the circulant matrix $\bS_h$ and
$c(h)=trace(\bOmega_h)/(n-p)$. The GCV
estimate of $h$ for  estimators of 
the form~(\ref{matrecon}) minimizes
\begin{equation}
\label{gcveqn}
\zeta(h)=\{\bz_1'(\bI_p-\bOmega_h)^2\bz_1 + [1+c(h)]^2\bz_2'\bz_2\},
\end{equation}
where $\bz=\bU'\by$, $\bz_1=\bU_1'\by$, $\bz_2=\bU_2'\by$.
\end{theorem}
\begin{IEEEproof}
See Appendix~\ref{appendix1}.
\end{IEEEproof}

The SVD of any $n\times p$ ($n>p$) matrix is generally expensive,
requiring computations on the order of at least 
$20p^3/3$~\citep{demmel97}. However, the complete SVD 
is unnecessary to calculate \eqref{gcveqn} and obtaining
$\bU_1'\by$  with $\bU_1$ as in Theorem~\ref{gcv.theorem} is
enough because $\bz_2'\bz_2$ can be computed from the identity
$\by'\by=\by'\bU\bU'\by
=\by'\bU_1\bU_1'\by+\by'\bU_2\bU_2'\by=\bz_1'\bz_1+\bz_2'\bz_2$. So we
devise a practical way to obtain $\bU_1'\by$.
Note that $\bU_1'\by =\bD_\bullet^{-1}\bV'\bK'\by$
where $\bD_\bullet$ is the diagonal matrix of the $p$ singular values of
$\bK$ with $\bV$ being the matrix of its right singular vectors.
Also, backprojection $\bK'\by$ is a necessary step in  BPF. Our
objective now is to efficiently compute $\bD_\bullet$
and $\bV'\bx$ for any 
vector $\bx$. We next derive some results on the eigendecomposition of
real symmetric circulant matrices.
\subsubsection{Spectral decomposition of circulant matrices}
\label{eigencircsec}
Let $\bC = \mbox{circ}(c_0,c_1,\ldots,c_{p-1})$ be a circulant
matrix with first row $\bc=(c_0,c_1,\ldots,c_{p-1})'$ and $\bgamma_{j,p} =  
(1,\omega_{j,p},\ldots,\omega_{j,p}^2,\omega_{j,p}^{p-1})'$ where
$\omega_{j,p}$ is the $j$th ($j=1,2,\ldots,p$) of the $p$ complex 
roots of unity. Then \citet{bellman60} shows that 
$d_j=\bc'\bgamma_{j,p}$ is the $j$th eigenvalue of $\bC$, with
corresponding eigenvector  $\bgamma_{j,p}$. Thus the
eigenvalues of any circulant matrix can be speedily
computed by using FFTs and scaling to equate the mean
to $c_0$. Also,  if $\bGamma_p$ is the matrix with $j$ column given by
$\bgamma_{j,p}$, then
$\bGamma_p'\bx$ is the forward Discrete Fourier Transform~(DFT) of
$\bx$ while  $\bGamma_p\bx$ is the inverse DFT of
$\bx$. However, these vectors are not necessarily real-valued and not
directly useful to us for finding $\bV$. So we derive further 
reductions for symmetric circulant matrices.
\begin{theorem}
\label{symmcirctheo}
Let $\bC$ be a $p\times p$ symmetric circulant matrix. Then the
eigenvalues of $\bC$ are all real and the 
spectral decomposition of $\bC = \bV\bD\bV'$ where, for even $p$,
$\bV=[\bone/\sqrt{p}, \bM_c,\pm\bone/\sqrt{p},\bM_s]$ with  $\bone=(1,1,\ldots,1)'$,
$\pm\bone=(1,-1,1,-1,\ldots,1,-1)'$, and $\bM_c$ and $\bM_s$ are $p\times
(p/2-1)$-matrices with $(j,k)$th element given by
$\sqrt{2/p}\cos{\left(2\pi k(j-1)/p\right)}$ and 
$\sqrt{2/p}\sin{\left(2\pi (p-k)(j-1)/p\right)}$,
respectively. Further, $\bD$ is the diagonal matrix of eigenvalues
with $k$th entry 
\begin{equation}
d_k=c_0+\sum_{j=1}^{p/2-1}c_j\cos\left(\frac{2\pi kj}p\right)
+c_{p/2}(-1)^k;\quad 0\leq k\leq p-1. 
\label{symmeigvalues}
\end{equation}
For odd $p$, the expression for the eigenvalues does not contain the 
last term. Also then, $\bV$ does not contain the column
vector $\pm\bone/\sqrt{p}$ and $\bM_s$, $\bM_s$ are $p\times (p-1)/2$-matrices. 
\end{theorem}
\begin{IEEEproof} See Appendix~\ref{appendix2}.\end{IEEEproof}
\begin{corollary}
\label{symmcalc}
For a real vector $\bx=(x_1,x_2,\ldots,x_p)'$, we have
\begin{enumerate} 
\item
\label{Vx}
$\balpha\doteq\bV'\bx$ can be computed
  directly from $\bbeta\doteq\bGamma'\bx$ because the
  first and $(p/2)$th (for $p$ even) elements of $\balpha$
  are the real parts of the corresponding elements of $\bbeta$. For
  $k=2,3,\ldots,[(p-1)/2]$, the $k$th element of $\balpha$ is the real
  part of the scaled sum of the $k$th and the $(p-k+2)$th elements of $\bbeta$
  while the $(p-k+2)$th element of $\balpha$ is the imaginary part of
  the scaled difference of the $k$th and the $(p-k+2)$th elements of 
  $\bbeta$. In both cases, the scaling factor is $\sqrt2$. Also, here  
  $[\xi]$ is the smallest integer that is no more than $\xi$. 
\item
\label{V'x} Let $\bpsi_1=(\sqrt{2}x_1,x_2,x_3,\ldots,\sqrt2^{{\mathcal 
    I}[p\mbox{ even}]}x_{[p/2]+1},\bzero')'$,   where $\bzero$ is a
vector of 0's and ${\mathcal I}[\cdot]$ is the indicator function. Also, let 
$\bpsi_2=(\bzero,x_{[p/2]+2},x_{[p/2]+3},\ldots,x_p)'$. Then each
    element of $\bV\bx$ is the  sum of the real and imaginary parts of
    the corresponding elements of $\bGamma'\bpsi_1/\sqrt{2}$ and
    $\bGamma'\bpsi_2/\sqrt{2}$, respectively. 
\end{enumerate}
\end{corollary}

\begin{IEEEproof} Part \ref{Vx} follows from the proof of
Theorem~\ref{symmcirctheo} while part \ref{V'x} follows by direct substitution.  
\end{IEEEproof}
Corollary~\ref{symmcalc} means that both $\bV'\bx$ and $\bV\bx$ can be
efficiently computed using FFTs. We now provide additional reductions on 2D
circulant matrices needed to calculate \eqref{gcveqn} for BPF.
\subsubsection{Spectral Decomposition of 2D Circulant Matrices}
\begin{definition}
\label{BCCB}
A 2D circulant matrix or,
alternatively, a block-circulant-circulant-block (BCCB) matrix is 
a $pq\times pq$-dimensional  matrix $\boldsymbol{\mathcal C}$  with
$p$ circulant blocks 
of $q$-dimensional circulant matrices. 
Thus, 
$\boldsymbol{\mathcal C}=\mbox{circ}(\bC^{(0)},\bC^{(1)},\ldots,\bC^{(p-1)})$, where each $\bC^{(i)}=\mbox{circ}(c_0^{(i)},c_1^{(i)},\ldots,c_{q-1}^{(i)})$.
\end{definition}
Note that a symmetric BCCB matrix necessarily has
symmetric blocks of symmetric circulant matrices. We now state a
result on the eigen-decomposition of such matrices. 
\begin{theorem}
\label{2dcirc.theo}
Let $\{\bgamma_{k,p}; k=1,2,\ldots,p\}$ and $\{\bgamma_{k,q};
k=1,2,\ldots,q\}$ be as in Section~\ref{eigencircsec}.
The $(k,j)$th eigenvalue of a  BCCB
matrix $\boldsymbol{\mathcal C}$ is 
$d_{k,j}=\sum_{l=0}^{p-1}\sum_{m=0}^{q-1}c_m^{(l)}\omega_{k,p}^l\omega_{j,q}^m$,
with eigenvector 
$\bgamma_{k,p}\otimes\bgamma_{j,q}$. Then the spectral decomposition
of $\boldsymbol{\mathcal
  C}=(\bGamma_p\otimes\bGamma_q)\bD(\bGamma_p\otimes\bGamma_q)'$
where $\bD$ is the diagonal matrix of eigenvalues
$\{d_{k,j};j=1,2,\ldots,q, k = 1, 2,\ldots,p\}$. 
\end{theorem}
\begin{IEEEproof}
The result follows by direct substitution and  the
fact that $\bgamma_{k,p}$ and $\bgamma_{k,q}$ are eigenvectors of 
$p\times p$ and $q\times q$ 1D circulant matrices,
respectively. \end{IEEEproof}

Theorem~\ref{2dcirc.theo} means that 2D FFTs can be used for
eigendecomposition of a BCCB matrix $\mbC$. More pertinently, 
the eigenvalues $d_{k,j}$ are scaled versions of the 2D FFT of
$\boldsymbol{\mathcal C}$, with scaling factor that equates the 
mean $d_{k,j}$ to the first element of
$\boldsymbol{\mathcal C}$. We now derive results for symmetric BCCB
matrices. 
\begin{corollary}
\label{2dsymmcirc.cor}
Let $\bV_p$ and $\bV_q$ be as in Theorem~\ref{symmcirctheo}.
Then the spectral decomposition of a symmetric BCCB matrix
$\boldsymbol{\mathcal C}$ is given by  $\boldsymbol{\mathcal
  C}=(\bV_p\otimes\bV_q)\bD(\bV_p\otimes\bV_q)'$,
with $\bD$ as in Theorem~\ref{2dcirc.theo}.
\end{corollary}
\begin{IEEEproof} Standard results on real symmetric matrices guarantee
  such a real-valued spectral decomposition. Replacing $\bGamma_p$ by    
$\bV_p$ and $\bGamma_q$ by $\bV_q$  in Theorem~\ref{2dcirc.theo}
yields the result.
\end{IEEEproof}

Corollary~\ref{2dsymmcirc.cor} means that for BCCB matrices, $\bV'\bx$ can
be computed for any $\bx$ using forward
FFTs. Hence, $\bU_1'\by$ of 
Theorem~\ref{gcv.theorem} is easily calculated in a
one-time calculation that can be used together with the
bandwidth-dependent parts of~(\ref{gcveqn}) to find the 
minimum. These latter calculations all involve linear
operations on the FFT results and can be speedily executed.
\subsection{Extension to Elliptically Symmetric Smoothing Kernels} 
\label{elliptical}
Most 2D FBP/BPF reconstruction filters are
radially symmetric. But the wider class of elliptically
symmetric kernels, such as the 2D Gaussian kernel with parameters $(h_1, h_2, \rho)$ 
\begin{equation*}
S_{h_1,h_2,\rho}(k,j)\propto\exp{\left\{-\frac1{1-\rho^2}\left(\frac{k^2}{h_1^2} + \frac{j^2}{h_2^2}   -2\rho \frac{kj}{h_1h_2}\right)\right\}},
\end{equation*}
provide greater flexibility because they allow for differential
smoothness along different directions and can better accommodate the natural
orientation of elongated structures.  However,  visually
selecting  optimal parameters for such kernels can be taxing because
of the larger set of parameters involved. Unlike FBP that uses 1D
filtering, BPF uses 2D filtering and so it is easy to incorporate such
kernels. Our development of Section~\ref{press} extends immediately, with $h$ in
\eqref{gcveqn} 
replaced by $h_1,h_2,\rho$ while optimizing \eqref{gcveqn}, making it 
possible to use elliptically symmetric smoothing kernels in BPF reconstruction.
\subsection{Overview of the GCV Bandwidth Selector}
We summarize here the steps of our method: 
\begin{enumerate}
\item {\em Corrected sinogram data.}  Get sinogram data $\by$ after 
  corrections for attenuation, scatter, randoms and so on.
  \item {\em Backprojection}. Backproject  $\by\rightarrow
    \tilde\blambda\doteq\bK'\by$.
  \item {\em Optimal bandwidth selection.}
    \label{opt} Apply the following
    steps: 
    \begin{enumerate}
    \item
      \label{step1}
      Obtain $\bD^2_\bullet\equiv\bD$ and (nominally)
      $\bV=\bV_p\otimes \bV_q$ for the approximately circulant
      $\bK'\bK$ following
      Theorem~\ref{2dcirc.theo} and
        Corollary~\ref{2dsymmcirc.cor}. Use forward FFTs to calculate
        $\bV'\tilde\blambda$ and get $\bz_1=\bU_1'\by =
        \bD^{-1}_\bullet\bV'\tilde\blambda$. Also, obtain $\bz_2'\bz_2
        = \by'\by-\bz_1'\bz_1$.
      \item
        \label{step2} For each $h$, obtain the eigenvalues, and hence
        $\bOmega_h$, of the circulant  
        smoothing matrix $\bS_h$, using
        Theorem~\ref{2dcirc.theo}. Calculate $\zeta(h)$ in
        \eqref{gcveqn}.  Then $h_{G}= 
        \argmin_h\zeta(h)$ is the GCV-estimated $h$.
    \end{enumerate}
  \item {\em Filtering.} \label{filt} The optimal GCV reconstruction is
    $\blambda^{h_G} = \bS_{h_{G}}(\bK'\bK)^{-1}\tilde\blambda$.
  \end{enumerate}
Our GCV selection method only needs the additional Step~\ref{opt}
beyond BPF reconstruction.  But Step~\ref{step1} is a one-time 
calculation, done by FFT, as also is Step~\ref{step2}, unless the
smoothing matrix is specified in the Fourier domain, in which case
$\bOmega_h$ is provided. Further, our algorithm outlined above details
the method for radially symmetric smoothing kernels. For elliptically-symmetric 
kernels as in Section~\ref{elliptical}, the $h$ is replaced by the
vector $(h_1,h_2,\rho)$ in Steps~\ref{step2} and \ref{filt}. 

A reviewer asked about $\bz_1$ and $\bz_2$. Our development
shows that $\bz_1$ to be a weighted version of the forward FFT of the
backprojected sinogram data, with the weights given by the square root
of the ramp filter. Further $\bz_2'\bz_2$ is the residual sum of
squares after removing the effect of the projection of
$\bz_1=\bU_1\by$ from the corrected sinogram data $\by$. Separately, as
also pointed out by the reviewer, the same matrices diagonalize the
circulant matrices $\bS_h$ and $\bK'\bK$ since their orders are the same.

\section{Performance Evaluations}
\label{experiments}
\subsection{Experimental Setup}
\label{setup}
\begin{figure*}[h]
\mbox{
\subfloat[Ground Truth]{
\label{hoff}
\hspace{-0.0225\textwidth}\includegraphics[width=0.22\textwidth]{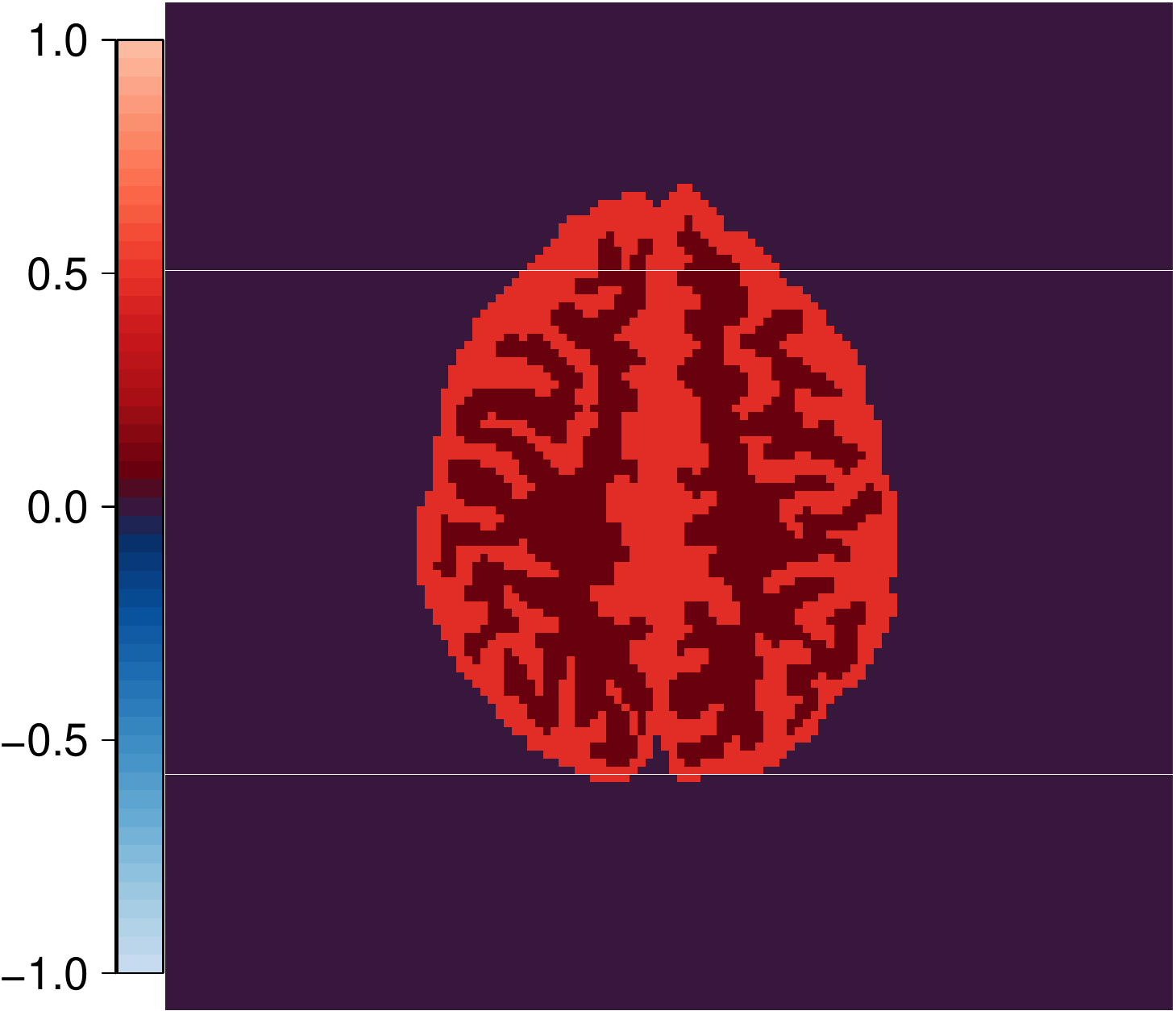}}
\subfloat[Gold standard FBP]{
\label{gold-radial}
\includegraphics[height=0.19\textwidth]{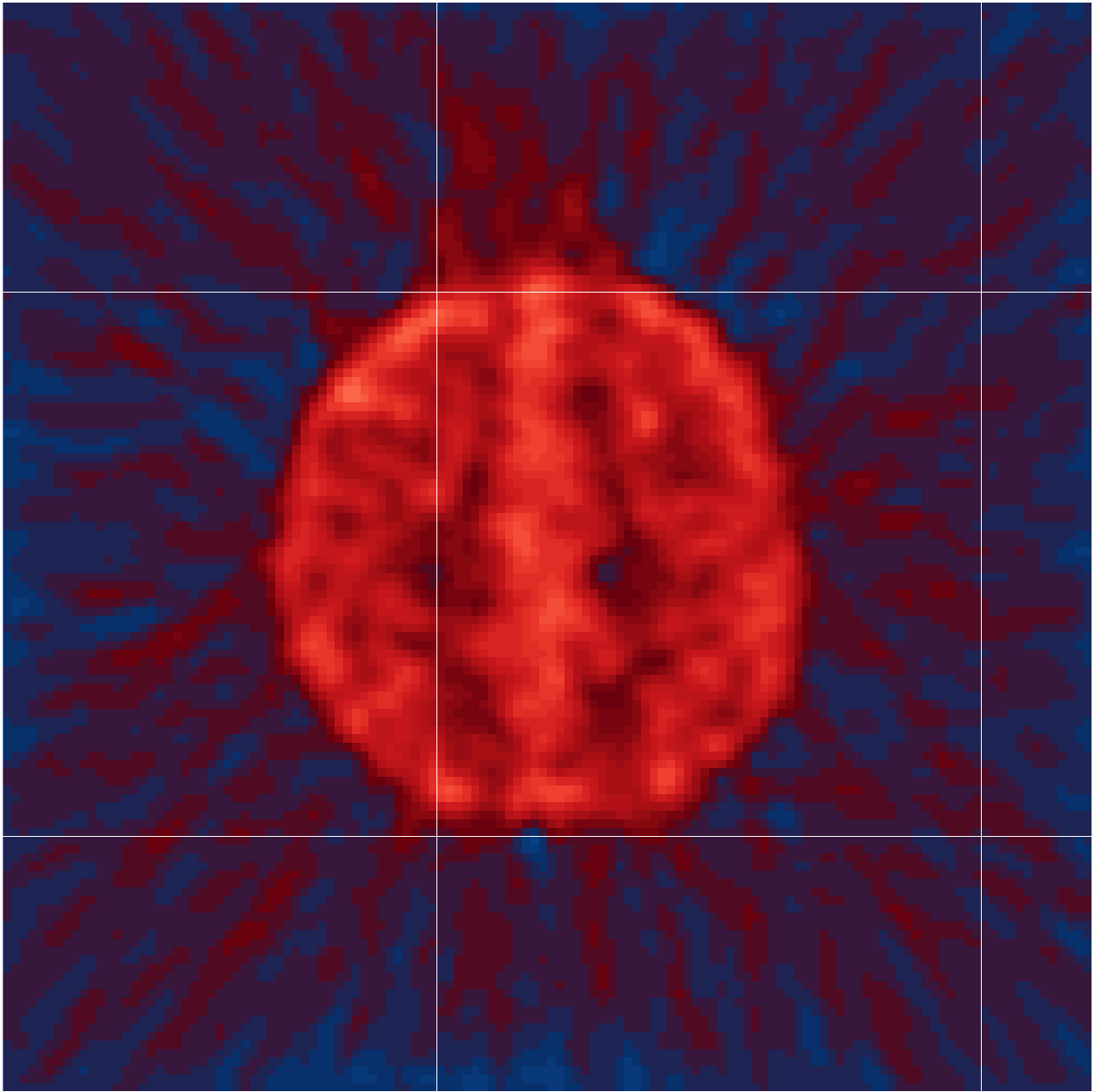}}
\subfloat[GCV-reconstructed FBP]{
\label{gcv-radial}
\includegraphics[width=0.19\textwidth]{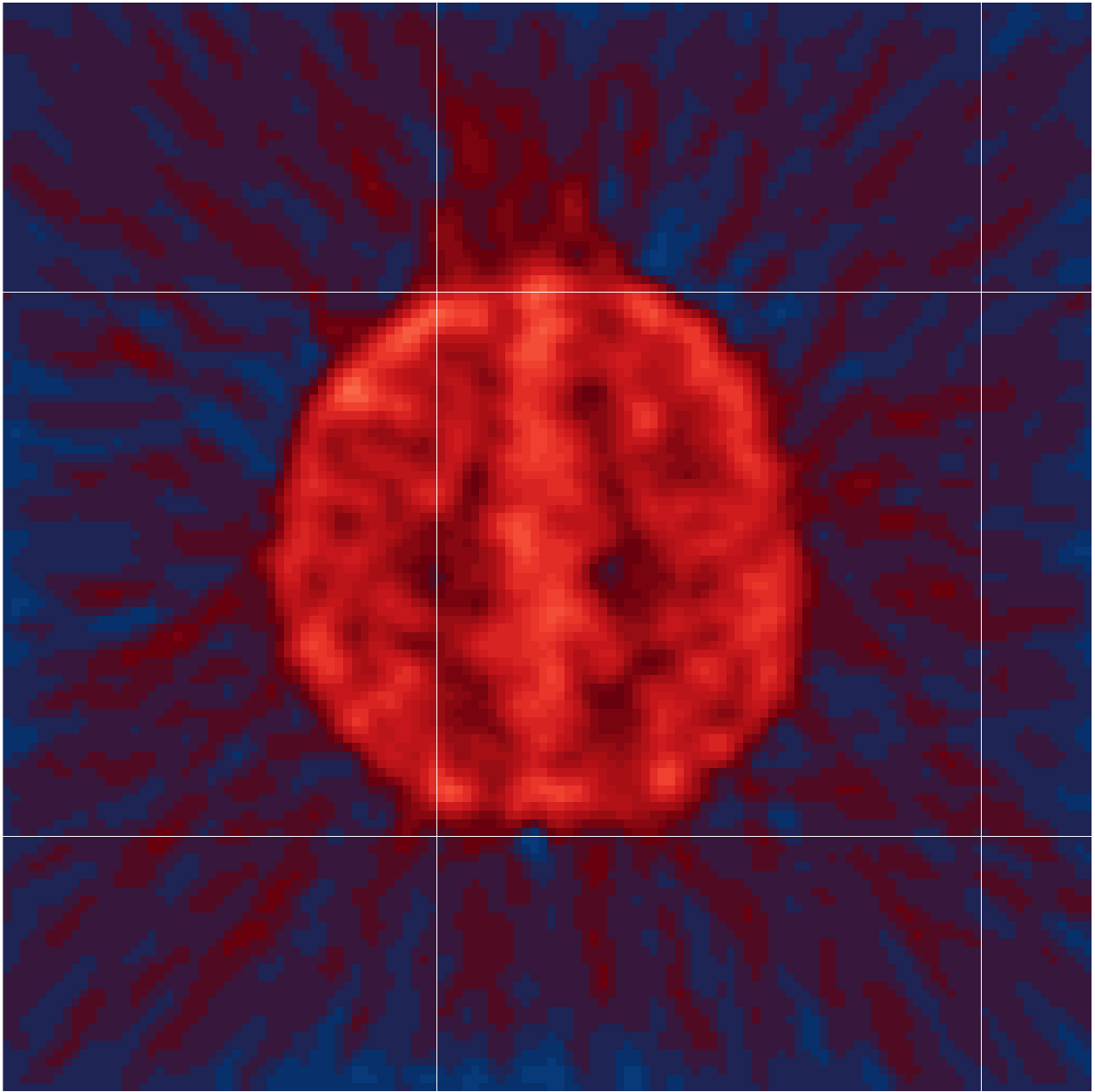}}
\subfloat[MRUP-reconstructed FBP]{
\label{mrup-radial}
\includegraphics[width=0.19\textwidth]{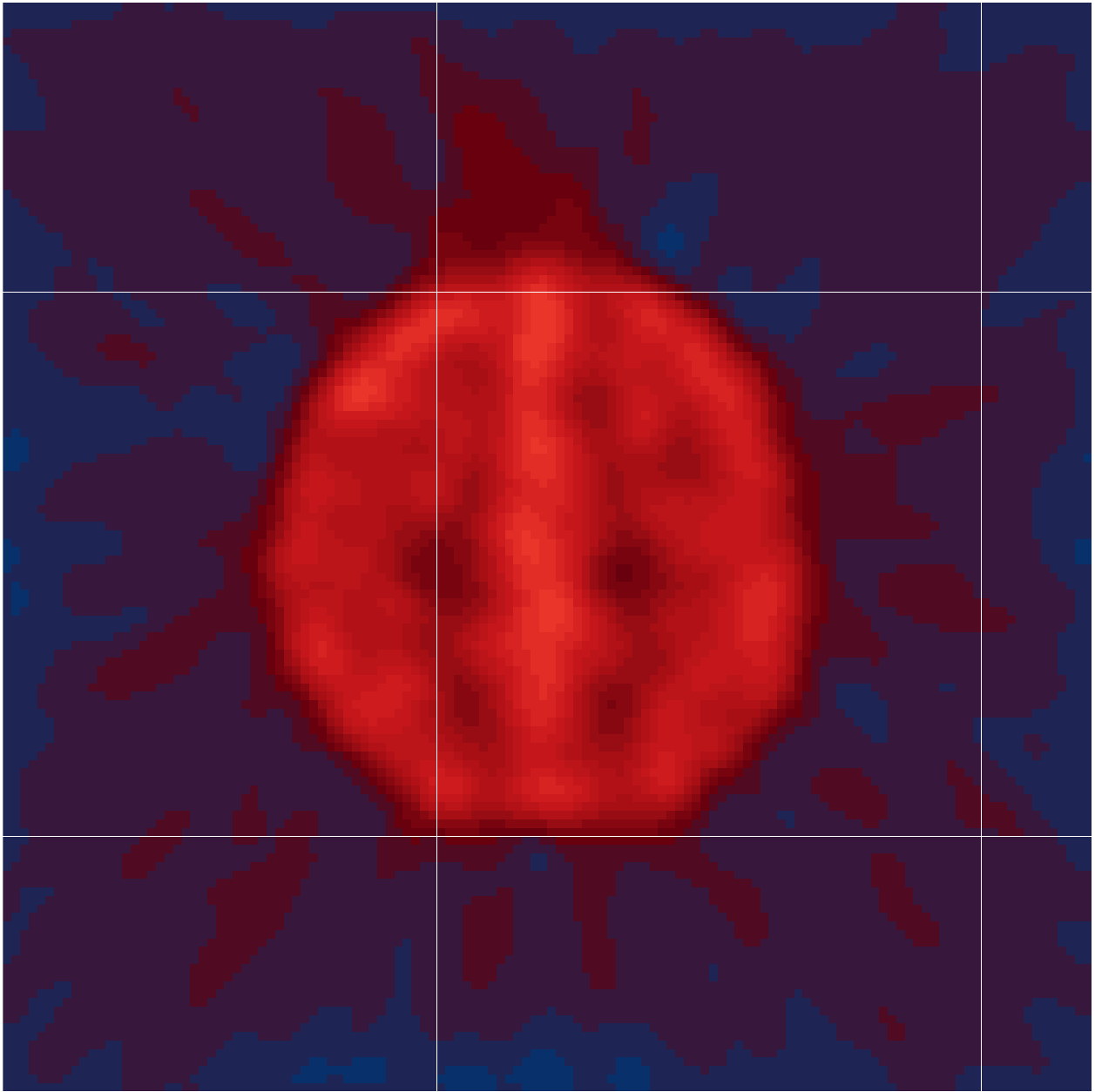}}
\subfloat[MRUE-reconstructed FBP]{
\label{mrue-radial}
\includegraphics[width=0.19\textwidth]{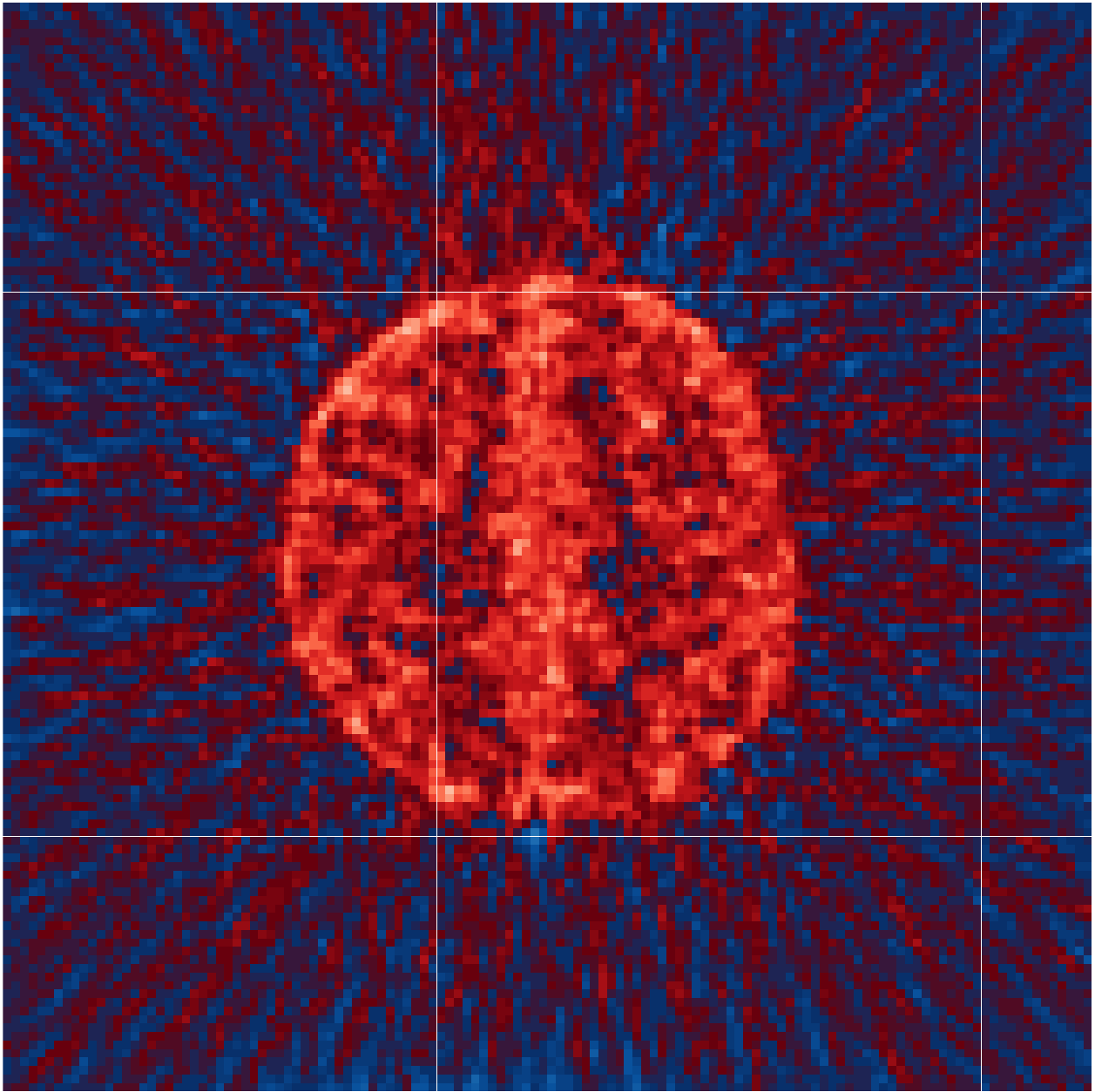}}
}
\caption{(a) The phantom, with colormap (ranging from [-1,1] to
  account for negative artifacts) that is used for all the
  reconstructed images in this article. 
  BPF reconstruction using the (b)
  optimal, (c) GCV- (d) MRUE- and (e) MRUP-selected bandwidths.}
\label{GCV}
\end{figure*}

The performance of our GCV approach 
was explored in a series of simulated but realistic 2D PET 
experiments. Our setup used the specifications and the sixth slice of
the digitized Hofmann~\citep{hoffmanetal90} phantom
(Figure~\ref{hoff}) on a discretized imaging domain having 
$128\times 128$ pixels of dimension 2.1 mm each. Our sinogram
domain had $128\times 320$ distance-angle bins~(LORs) of size
$2.1\mbox{mm}\times\pi/320\mbox{ radians}$. 
Pseudo-random Poisson realizations were simulated in the sinogram
domain with mean intensity given by the corresponding discretized Radon 
transform of the phantom. The total expected counts $\Lambda$ varied over 9 distinct  equi-spaced (on a $\log_2$
scale) values between $10^4$ and $10^6$ counts.
Therefore, $\Lambda$ ranged from the very low~(about 0.61
counts per pixel) to the moderately high~(61 counts per
pixel) and matched the range of values typically seen in
individual scans in dynamic 2D PET studies~\citep{phelpsetal79}. 
Our first set of evaluations used a radially symmetric Gaussian
kernel $\bS_h$ with FWHM $h$. Subsequent evaluations used
elliptically symmetric Gaussian kernels with parameters 
$h_1,h_2,\rho$. We use ``BPFe'' to denote BPF reconstructions with
elliptically symmetric kernels and ``GCVe'' to denote GCV estimated
parameters in these settings. We also evaluated performance in
applying reducing negative artifacts as per
\citet{osullivanetal93}  -- we add ``+'' in the nomenclature to
denote this additional postprocessing step. For each simulated
sinogram dataset, we obtained the optimal BPF, 
BPFe, BPF+ and BPFe+ reconstructions and the corresponding optimal 
bandwidths as follows: for the BPF reconstruction
$\hat\blambda^{h}$ using a radially symmetric Gaussian filter with
FWHM $h$, we calculated the Root Mean Squared Error (RMSE) $\mR_{h}
= \parallel\!\hat\blambda^h - 
\blambda\!\parallel/\sqrt{p} $, with $\blambda$ the true source
distribution (the Hoffman
\citep{hoffmanetal90} phantom). The $ h _O$ corresponding to the BPF
reconstruction that minimizes the RMSE ({\em i.e.}, the $h_O$ such
that $\mR_{h_O} \leq \mR_{h}$ for all $h\neq h_{O}$) is our {\em true
  optimal} FWHM for the simulated sinogram dataset. Similar optimal
FWHM parameters and gold standard reconstructions were obtained for
BPFe, BPF+ and BPFe+ reconstructions (note that BPFe and BPFe+ have
trivariate smoothing parameters.) We evaluated
performance of reconstructions obtained using our GCV-estimated
procedure ($h_G$) in terms of the RMSE and compared them (for BPF and
BPF+ reconstructions)  with corresponding calculations  obtained using
the MRUE- and MRUP-estimated~\citep{pawitanandosullivan93,osullivanandpawitan96}
bandwidths $h_E$ and $h_PE$, respectively. We also evaluated performance
of each reconstruction  in terms of
its RMSE efficiency relative to the gold standard reconstruction
obtained using $h_O$, that is, we calculated RMSE efficiency for a BPF
reconstruction with filter resolution $h$ as $\mR_h/\mR_{h_O}$.
Corresponding evaluations were done for BPFe, BPF+ and BPFe+
reconstructions, with smoothing parameters in BPFe and BPFe+ optimized
over trivariate sets. Reducing negativity artifacts does not involve
choosing a $h$ beyond that chosen for 
BPF or BPFe; however the optimal BPF+ or BPFe+ bandwidths
may be  different from the optimal BPF and BPFe ones. 
We simulated 1000 sinogram datasets and evaluated
reconstruction performance using the different methods.
\subsection{Results}
\subsubsection{Illustrative Examples} 
We first illustrate performance on a sample simulated sinogram
realization with $\Lambda=10^5$.
\paragraph{BPF reconstruction} 
Figure~\ref{gold-radial} provides the ``gold standard'' BPF/FBP reconstruction
\label{results:radial}
\begin{table}[h]
  \caption{Bandwidths and RMSEs obtained with BPF reconstructions
    using bandwidths from different selection methods.}
  \label{tabs1}
  \begin{center}
  \begin{tabular}{ccccc} \\ \hline
    {\bf Estimation Method} & Gold Standard & GCV & MRUP & MRUE \\  \hline
    Optimal Bandwidth & 3.183 & 3.279 & 15.32 & 1.397 \\ 
    RMSE ($\times 10^{-5}$) & $8.376$ & $8.378$ & $10.32$ & $10.83$ \\ \hline
  \end{tabular}
  \end{center}
\end{table}
obtained  using $h_O$. BPF reconstructions obtained using $h_G$, $h_P$
and $h_E$ are in Figures~\ref{gcv-radial},~\ref{mrup-radial}
and~\ref{mrue-radial}, respectively. Table~\ref{tabs1} provides
the estimated bandwidths and numerically summarizes performance in
terms of the RMSEs. Performance using MRUE~($h_E=1.397$ pixels, RMSE$=
1.083\times 10^{-4}$) and  MRUP ($h_P=15.322$ pixels, RMSE$= 
1.032\times 10^{-4}$) bandwidths is not satisfactory, with the 
methods considerably under- and over-estimating the 
bandwidths, respectively. (Following~\citet{pawitanandosullivan93},
$h_P$  is specified in the 1D filtering domain of the projection
distances, and is not directly comparable in numerical value to the 2D
filter bandwidth). On the other hand, GCV 
 ($h_G= 3.279$ pixels, RMSE $=8.378\times 10^{-5}$) tracks the
optimal value ($h_O= 3.183$ pixels, RMSE $=8.376\times 10^{-5}$)
very closely, both in terms of bandwidth selection and reconstruction
ability. 

\begin{figure}[h]
    \centering
    \mbox{
      \subfloat[BPFe Gold standard]{
        \label{gold-elliptical}
        \includegraphics[height=0.19\textwidth]{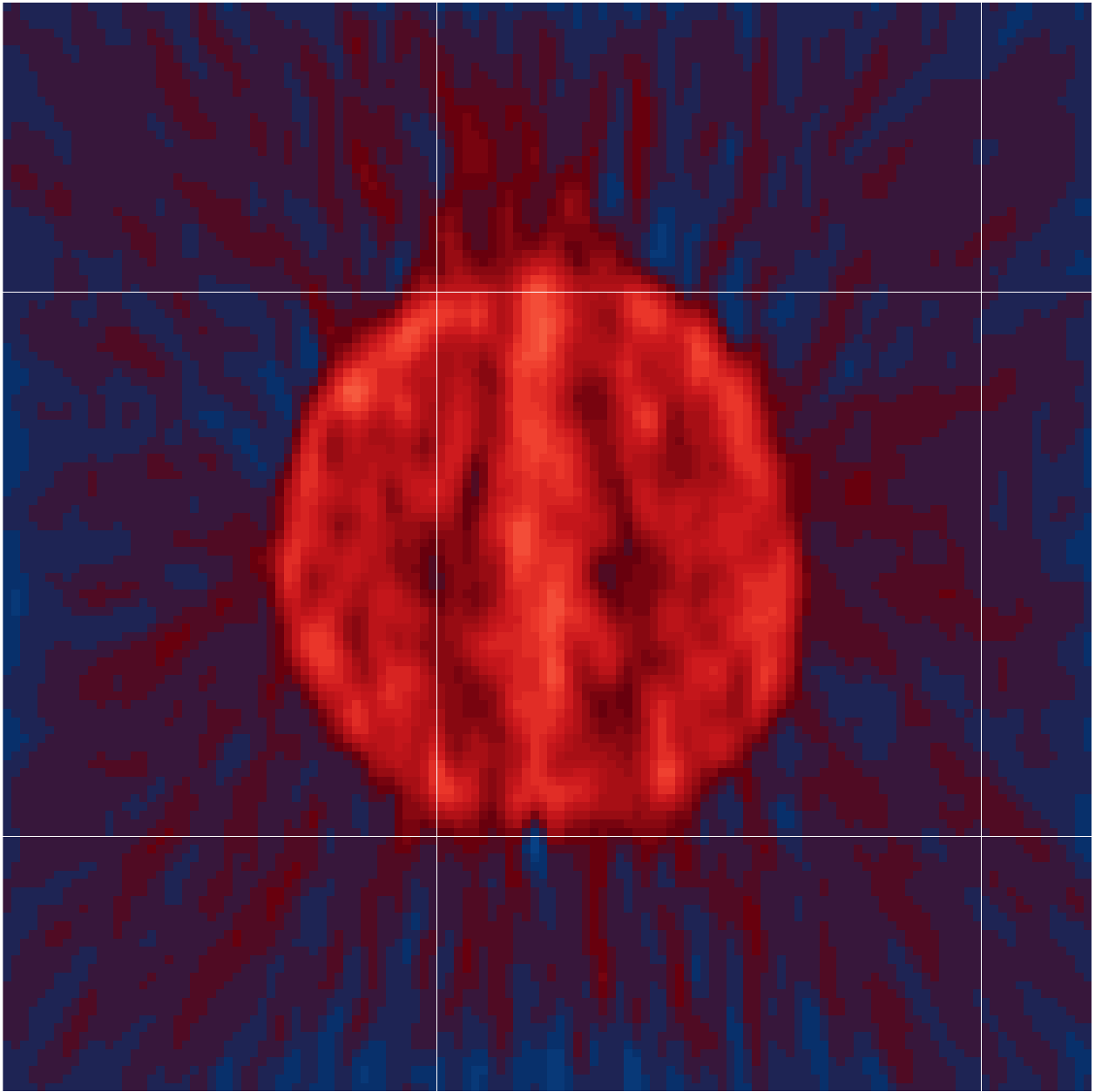}}
      \subfloat[GCVe reconstruction]{
        \label{gcv-elliptical}
        \includegraphics[width=0.19\textwidth]{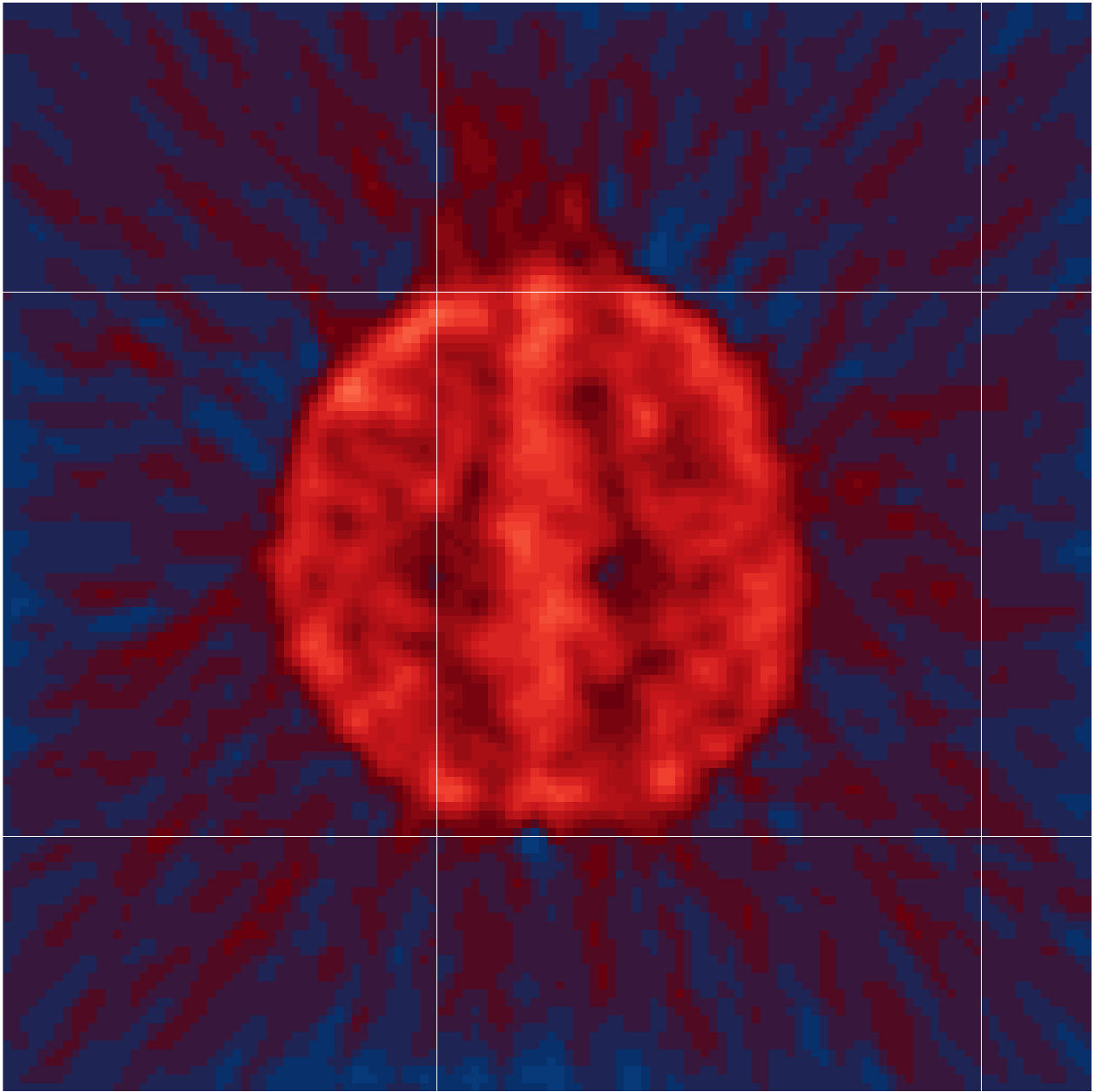}}}
\subfloat[Reconstruction performance]{
  \label{tabs2}
      \raisebox{1.25\height}{\begin{tabular}{ccc} \\ \hline
    {\bf Estimation Method} & Optimal Parameters &   RMSE    \\
 & $(h_1,h_2,\rho)$ & ($\times  10^{-5}$)\\ \hline 
    Gold Standard & (5.143, 2.488, -0.122) & 8.289    \\
    GCVe-estimated & (3.249, 3.058, 0.170) & $8.378$ \\ \hline
  \end{tabular}
}
}
    \caption{BPFe reconstructions with the (a) optimal and (b)
      GCVe-estimated smoothing parameters along with (c) summary of
      reconstruction performance.}
    \label{fig:elliptical}
\end{figure}

\paragraph{BPFe reconstruction}
\label{results:elliptical}
We next illustrate GCVe's performance 
in  choosing optimal parameters for BPFe reconstructions.
Figure~\ref{fig:elliptical} shows reconstructions obtained using the
{\em true optimal} and the GCVe-selected  parameters. The
\begin{figure*}
\mbox{
\subfloat[BPF+ Gold standard]{
\label{gold-bpfp}
\includegraphics[width=0.19\textwidth]{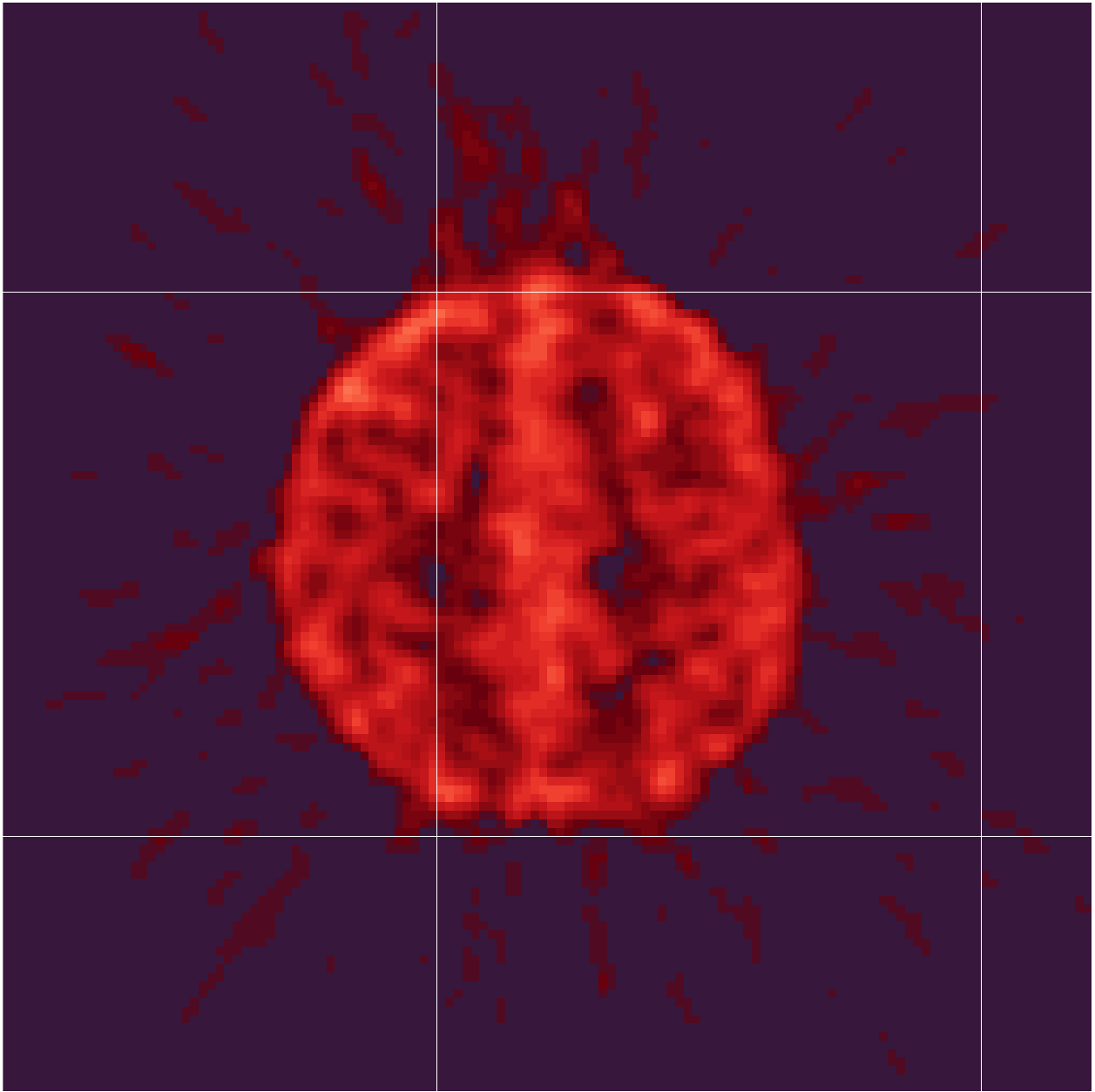}}
\subfloat[GCV+]{
\label{gcv-bpfp}
\includegraphics[height=0.19\textwidth]{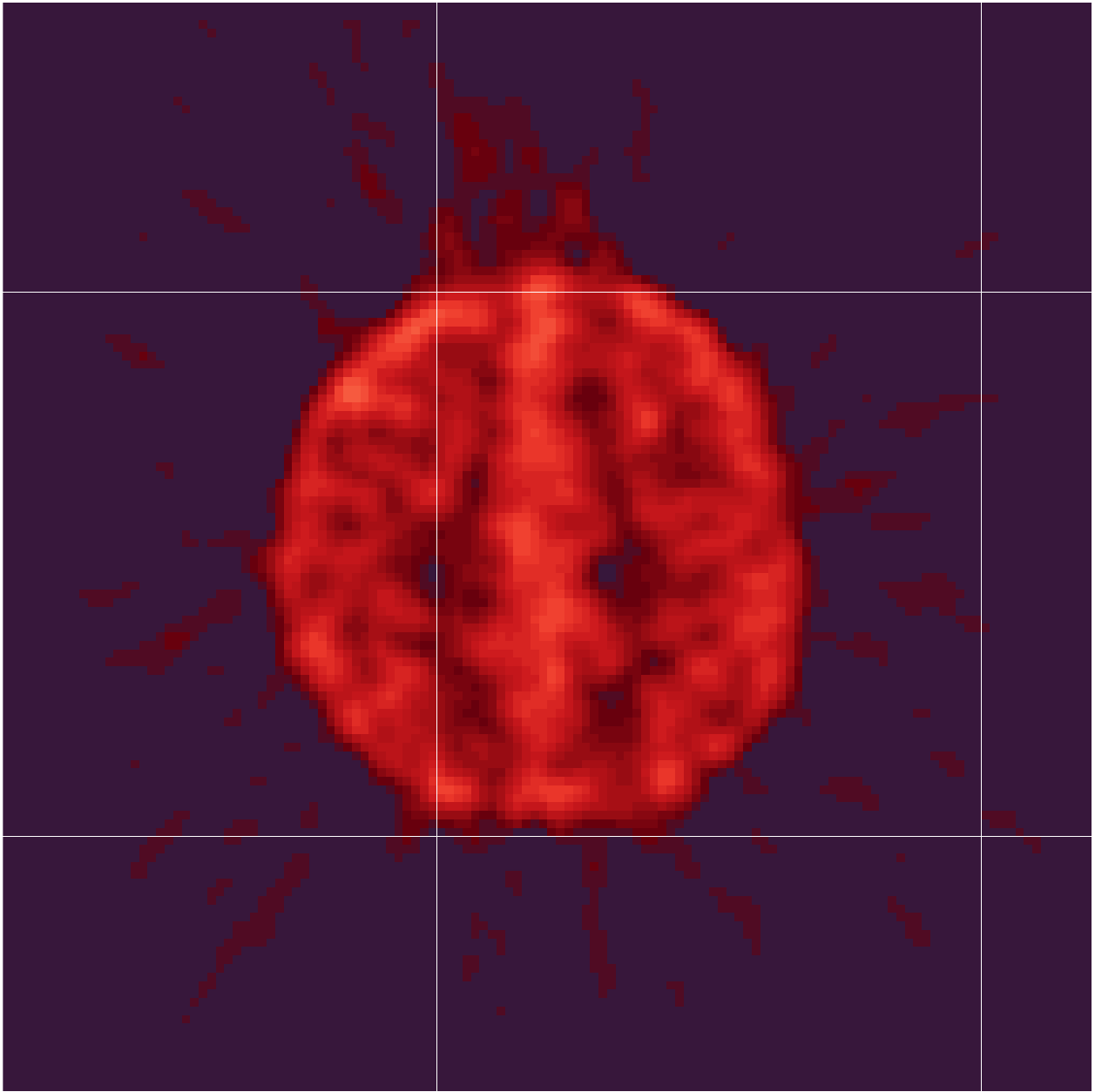}}
\subfloat[MRUP+]{
\label{mrup-bpfp}
\includegraphics[width=0.19\textwidth]{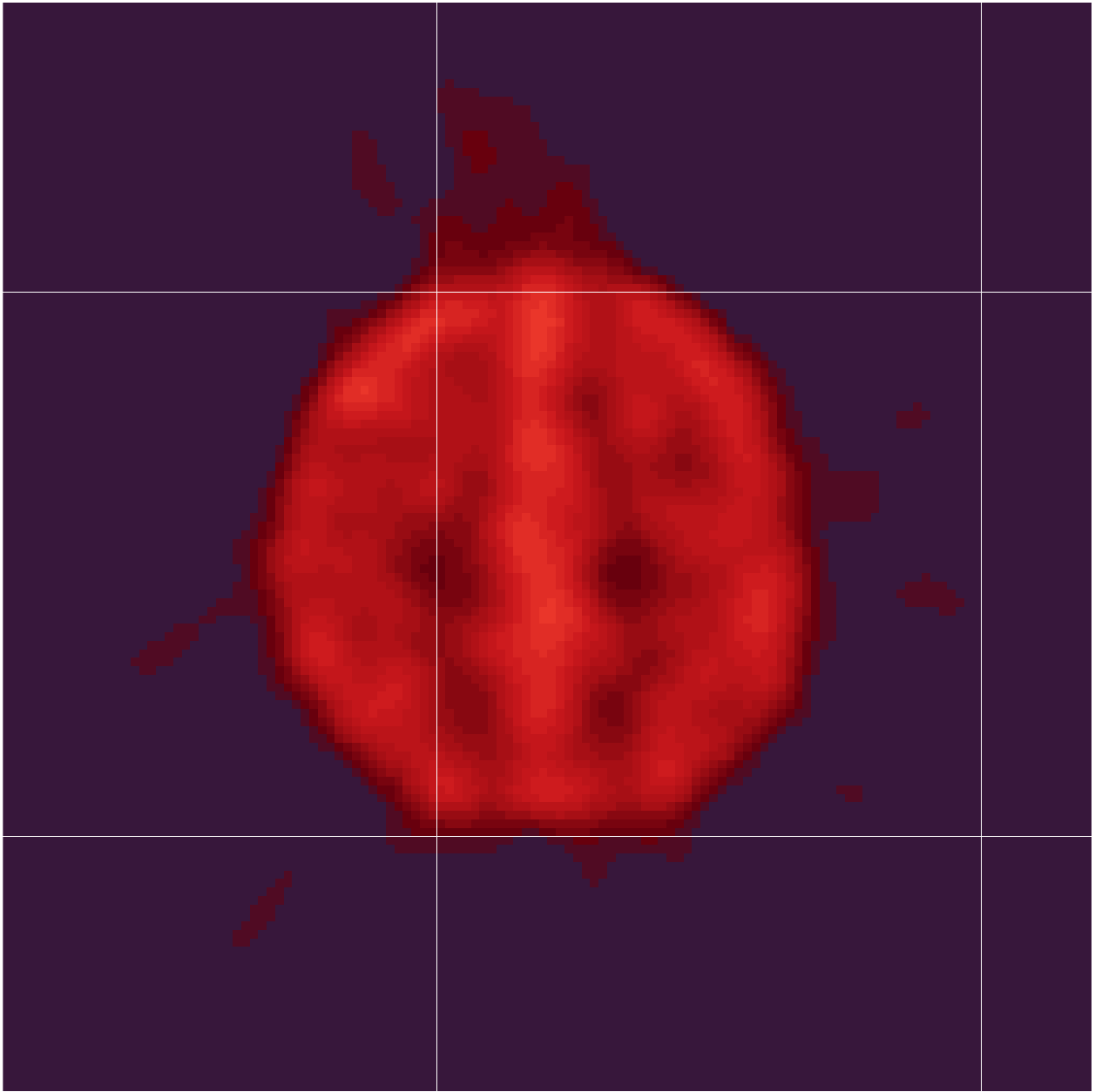}}
\subfloat[BPFe+ Gold standard]{
\label{gold-bpfp-noradial}
\includegraphics[width=0.19\textwidth]{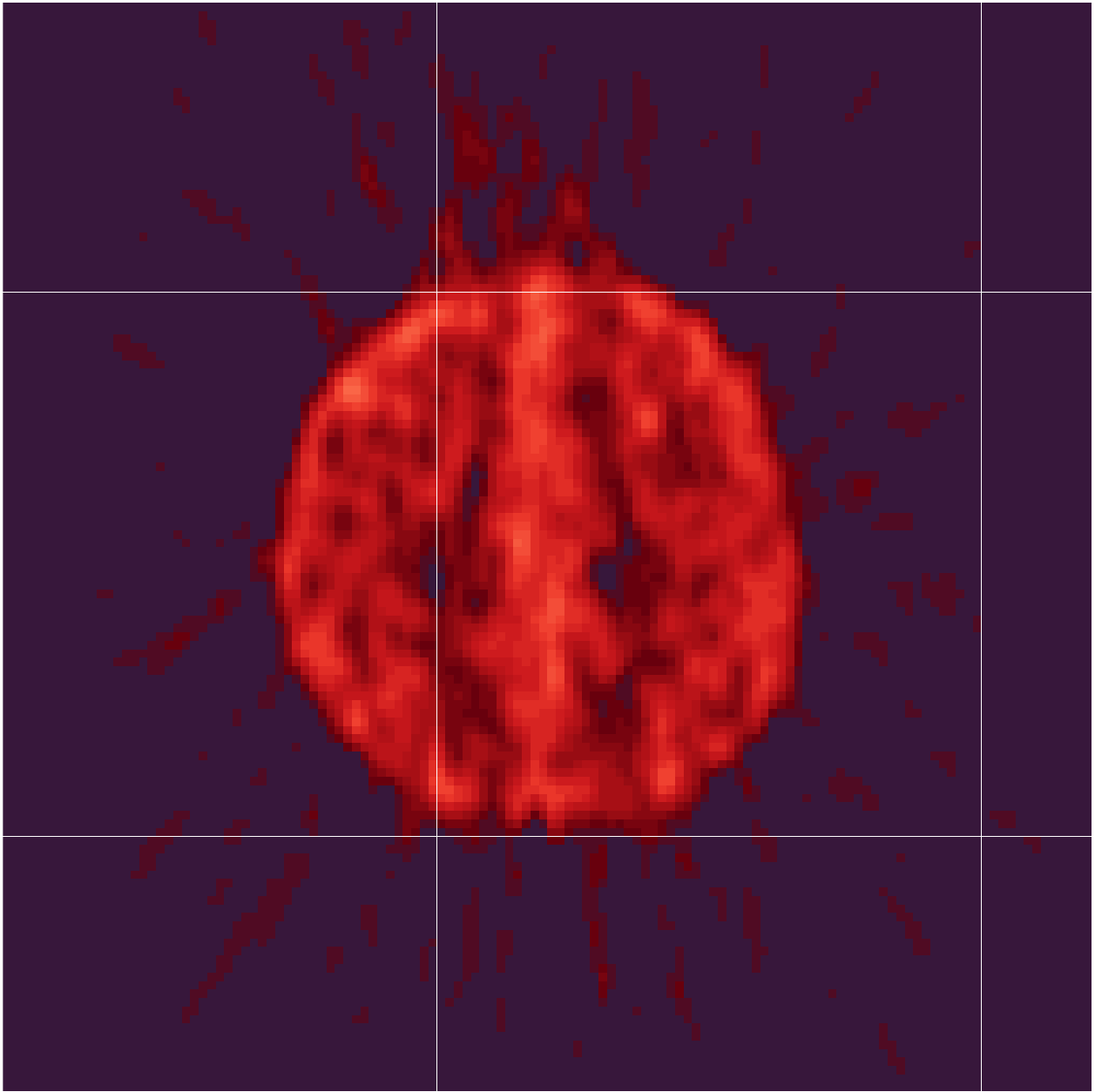}}
\subfloat[GCVe+]{
\label{gcv-bpfp-noradial}
\includegraphics[width=0.19\textwidth]{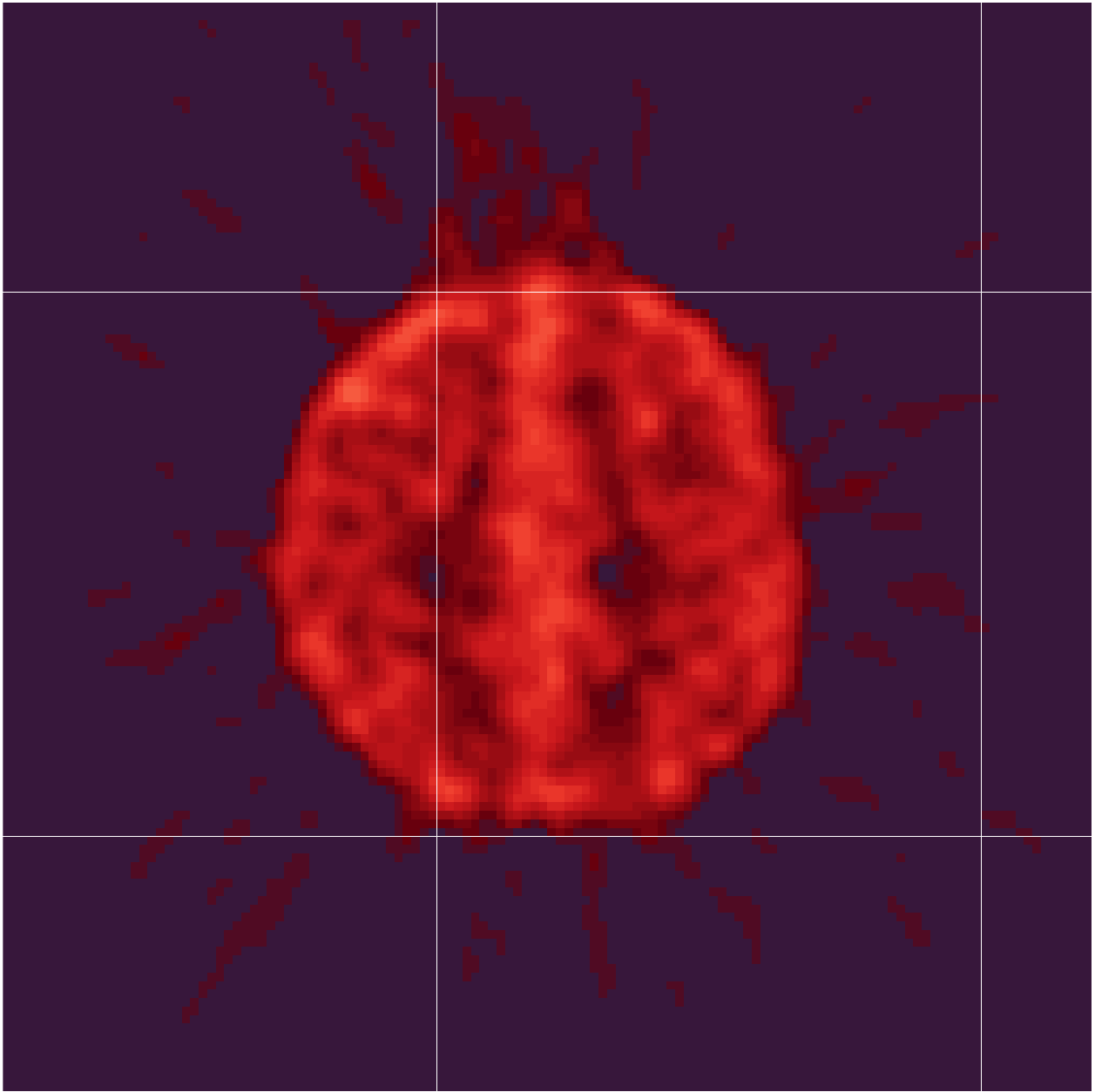}}
}
\caption{(a-c) BPF+ reconstructions using the (a) optimal, (b) GCV-
  and (c) MRUP-estimated bandwidths. (d, e) BPFe+ reconstructions using the
  (d) optimal and (e) GCVe-estimated parameters.}
\label{FBPP}
\end{figure*}
reconstruction RMSE (Figure~\ref{tabs2}) of $8.367\times 10^{-5}$ 
obtained using BPFe with GCVe even marginally outperforms BPF
reconstruction using $h_O$ and compares favorably with
the  true optimal RMSE of $8.289\times 10^{-5}$.  The illustration
shows some undersmoothing with GCVe and scope for improved parameter 
estimates, but the wider class of elliptically symmetric kernels throws
opens the possibility for further improvements to
reconstruction.
\paragraph{Reconstructions with reduced negative artifacts}
We also explored performance  in BPF+ and
BPFe+ reconstructions, having gold standards as per
Figure~\ref{gold-bpfp} ($h_{O+} = 2.766$; RMSE =  
$7.873\times 10^{-5}$) and Figure~\ref{gold-bpfp-noradial}
($h_1=3.818, h_2=2.216, \rho=-0.035$; RMSE = $7.831\times 10^{-5}$),
respectively. For BPF+, the GCV-estimated bandwidth of $h=3.279$
yields the reconstruction of Figure~\ref{gcv-bpfp} (RMSE$=7.976\times
10^{-5}$) while MRUP provides the BPF+ reconstruction 
of Figure~\ref{mrup-bpfp} (RMSE=$1.008\times 10^{-4}$). For brevity
of display, we forego discussing BPF+ reconstructions done with MRUE-selected
bandwidths, noting simply that they also improve over BPF under BPF+
(with RMSE=$8.646\times 10^{-5}$ in this example) but that improvement
falls far short of that obtained using
GCV. Figure~\ref{gcv-bpfp-noradial} also shows improvement of BPFe+
over BPFe (RMSE = $7.968\times 10^{-5}$) when negative artifacts are eliminated
using~\citep{osullivanetal93}, but the improvement 
is very marginal.  Note that \citet{osullivanetal93}
reduces negative artifacts using a radially symmetric filter -- an
alternative approach that allows for greater flexibility in 
smoothing out negative values may be more appropriate.
\subsubsection{Large-scale Simulation Study}
\label{simulation}
\begin{figure*}
  \centering
  \mbox{
    \subfloat[]{\label{2d-rmse-gcv}\includegraphics[width=\textwidth]{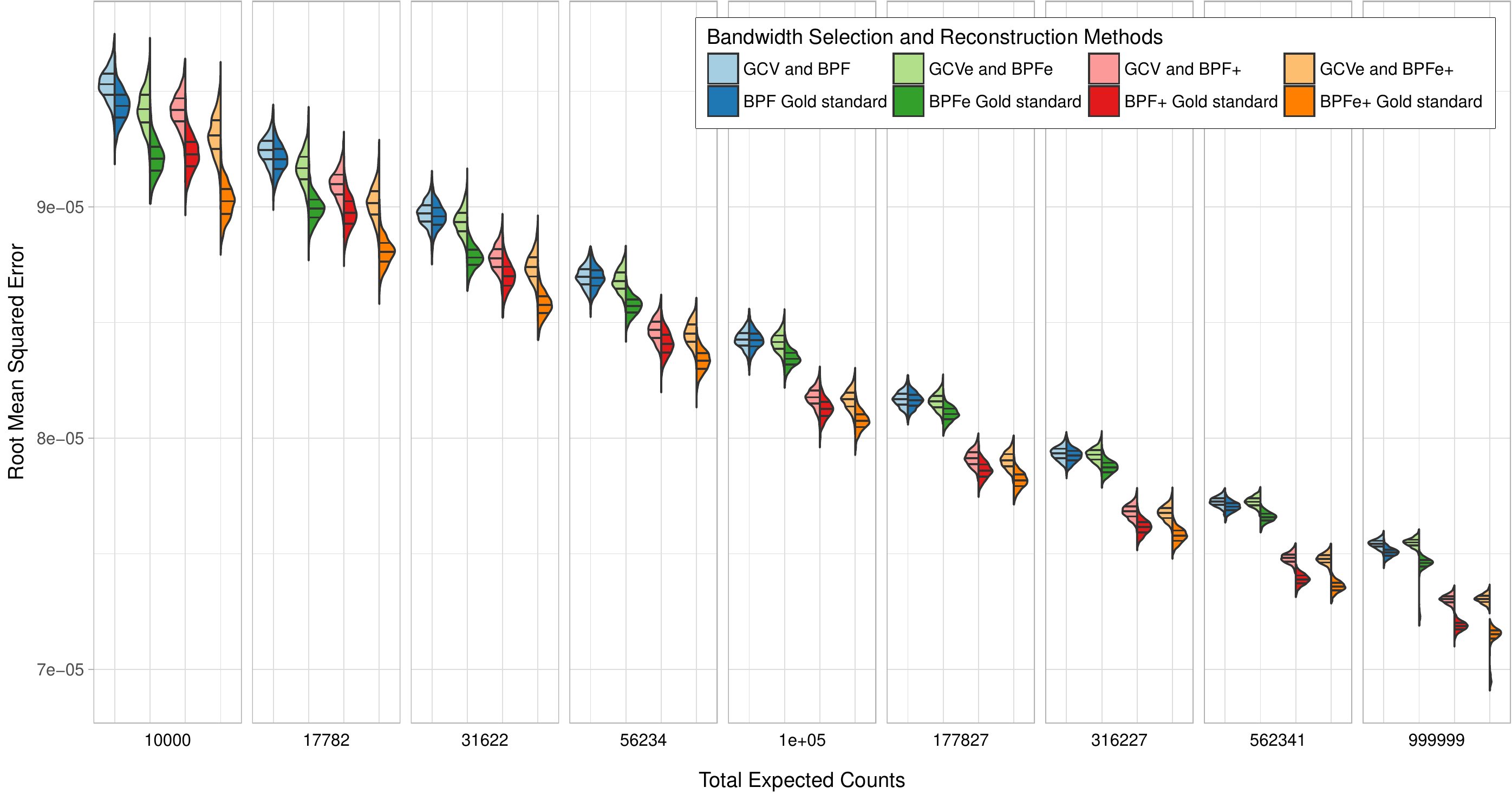}}
  }
  \mbox{\subfloat[]{\label{2d-rmse-mrue}\includegraphics[width=0.5\textwidth]{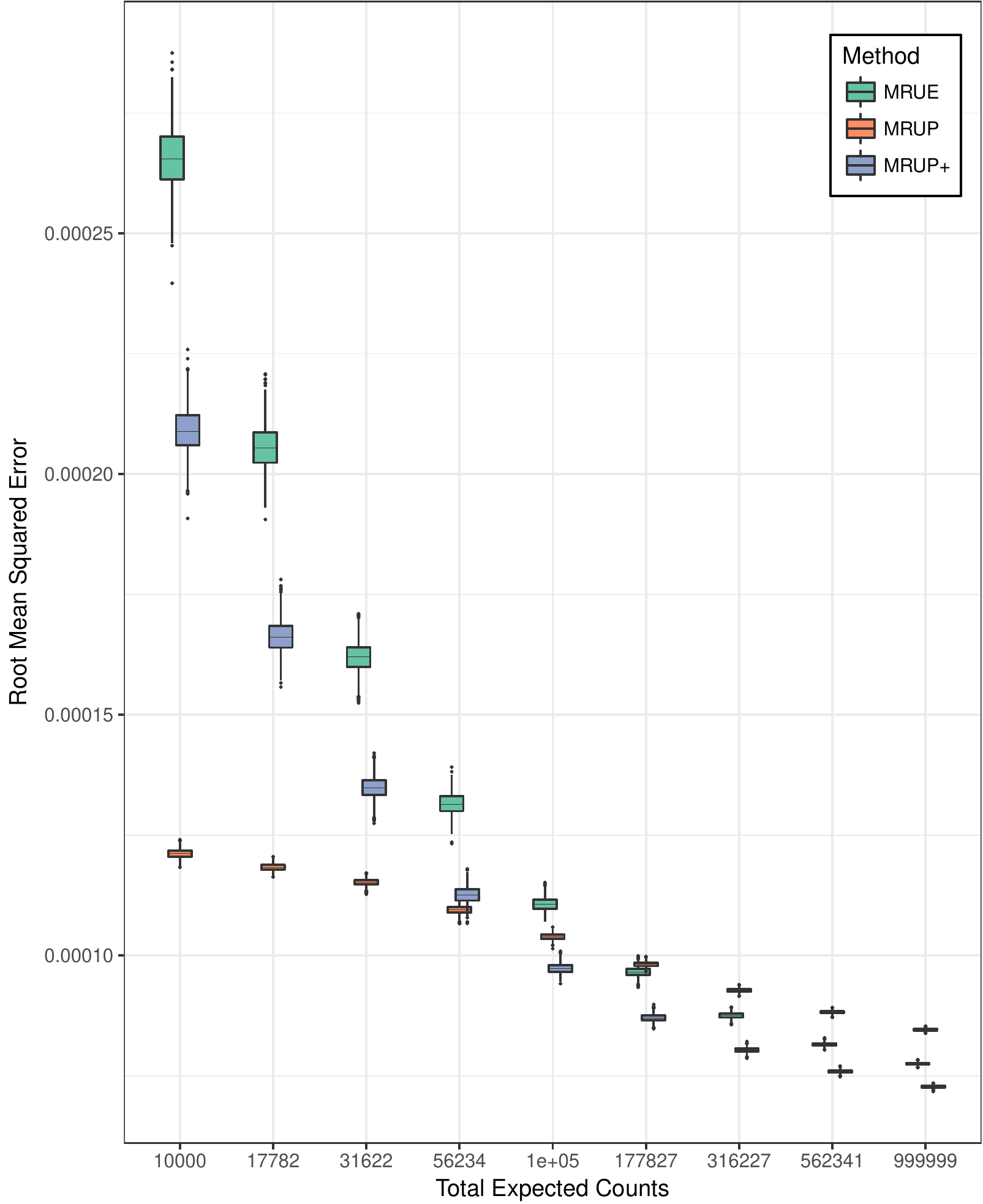}}
    \subfloat[]{\label{2d-rmse-effs}\includegraphics[width=0.5\textwidth]{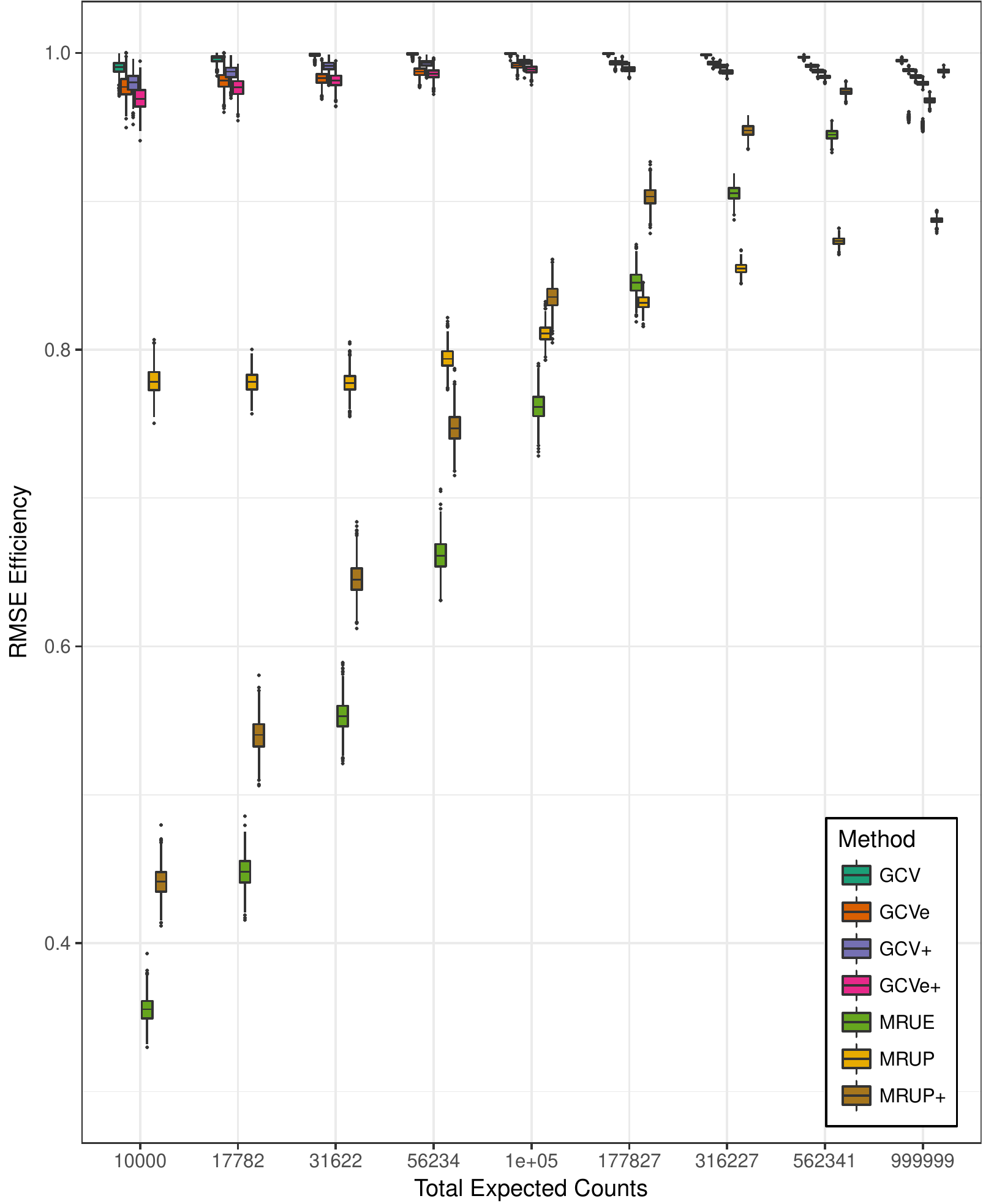}}
  }
\caption{(a) Split violin plot distributions against total expected counts
  ($\Lambda$) of the 1000 RMSEs  for reconstructions using
  GCV-estimated (left violin
  lobe) and  the corresponding gold standard (right
  lobe) reconstruction. For each $\Lambda$-value, violins are in the order of
  BPF, BPFe, BPF+ and BPFe+ reconstructions. Bars on each split violin
  display the upper, median and lower quartiles of the RMSEs. (b)RMSEs of MRUE, MRUP and MRUP+ reconstructions
  and (c) RMSEs of all reconstructions relative to the
  corresponding gold standard for different total expected counts
  using different bandwidth selection and reconstruction methods.} 
\end{figure*}

We report results of our large-scale simulation study on the performance (in
terms of RMSE) of the different bandwidth selection and reconstruction 
methods  and their distribution for different values of
$\Lambda$. Reconstructions using MRU bandwidths have 
RMSEs substantially higher than those using the optimal or
GCV-estimated bandwidths (Figure~\ref{2d-rmse-gcv}) and certainly for
lower values of $\Lambda$, so we display performance of these
estimators separately in Figure~\ref{2d-rmse-mrue} in order to attain
finer granularity for displays involving our methods.
Figure~\ref{2d-rmse-gcv} displays RMSEs of BPF, BPFe, BPF+ and BPFe+
reconstructions obtained with GCV and the corresponding true optimal
bandwidth parameters. 
The BPFe reconstructions using the GCVe-estimated
bandwidths have similar, if not lower RMSEs, to those obtained with the
gold standard BPF reconstructions. Reducing negative artifacts \citet{osullivanetal93} improves the quality of BPF or BPFe reconstructions that
is more substantial at higher $\Lambda$-values. But the
improvement with using GCVe-estimated elliptically symmetric filters
over GCV-estimated radially symmetric filters 
tapers off at higher total expected counts. However, the optimal BPFe
estimator improves 
reconstruction quality in terms of having lower RMSEs over the gold
standard BPF reconstructions. Thus, the performance of
GCVe-estimated reconstruction relative to the gold standard BPFe is
not as strong as that of the GCV-estimated reconstruction relative
to the BPF gold standard. This observation is also
supported by the relative RMSE efficiency displays in
Figure~\ref{2d-rmse-effs}. This may be because, as
per the table in Figure~\ref{tabs2} of our illustrative example, the bandwidth
parameter sets are quite different than the true optimal BPFe
parameters. 
Nevertheless, Figure~\ref{2d-rmse-gcv} shows that any of the GCV methods
out-performs the MRUE methods, especially at low total expected
counts, both in terms of raw RMSE (Figure~\ref{2d-rmse-mrue}) and
relative RMSE efficiency 
(Figure~\ref{2d-rmse-effs}). Indeed, the relative RMSE efficiencies are almost always above 0.95
for the GCV methods.  However, the MRU reconstructions are rather
poor, especially at lower values of $\Lambda$. The MRUP results
reported here are a bit more pessimistic than those over limited
$\Lambda$ reported in 
\citet{pawitanandosullivan93} and \citet{osullivanandpawitan96}. Interestingly 
and contrary to their results, for larger (but not smaller) values of
$\Lambda$, MRUE  outperforms MRUP: comparison with their computer code
indicates that the optimal bandwidths are often attained outside their
chosen ranges for several cases. Reducing negative artifacts as per
\citet{osullivanetal93} improves MRUE reconstructions slightly -- we
omit these RMSEs in 
Figures~\ref{2d-rmse-mrue} and \ref{2d-rmse-effs} for clarity of
display. The methods of \citet{osullivanetal93} degrades MRUP
reconstructions for lower $\Lambda$-values but with increasing
$\Lambda$, MRUP+ generally performs the best among all MRUE
estimates. The rate of efficiency of  reconstructions with increasing
$\Lambda$ obtained using GCV-selected bandwidths is lower
than either MRUE or MRUP, but the implications are unclear, given
its superior performance at all $\Lambda$.  
The results point to the ability of 
the GCV-estimated bandwidths in obtaining improved reconstructions in
situations with low and high expected total emissions.

\subsection{Some Theoretical Analysis of the GCV Selector}
We now discuss some theoretical properties of the  
reconstructions obtained using the GCV-selected bandwidth. Because our
primary setting for investigating bandwidth selection in this article is PET and
because of the additional complication provided by the Poisson 
distribution of the emissions, our investigation is in the context of
an idealized emission tomography experiment. Suppose that we have
$\by$ realized from an  inhomogeneous Poisson Process with $\E(\by) = \bmu =
\bK\blambda$, and $\Lambda=\sum_{d,\theta}\mu_{d,\theta}$. Our interest
is in estimating $\bff=\blambda/\Lambda$, 
for which we propose the  estimator $\hat\bff^h =
\hat\blambda^h/\hat\Lambda$. As in \citet{osullivanandpawitan96}, 
define the loss function in the estimation and prediction domains to be
$L_e(\hat\bff^h,\bff) = \parallel\!\hat\bff^h - \bff\!\parallel^2$ and 
$L_P(\hat\bmu^h,\bmu) = \Lambda^{-2}\parallel\!\hat\bmu^h - \bmu\!\parallel^2$,
respectively, with corresponding risk functions as ${\mathcal 
  R}_e(\hat\bff^h,\bff)$ and  ${\mathcal R}_P(\hat\bmu^h,\bmu)$.  
From the SVD of $\bK=\bU_1\bD_\bullet\bV$ (where $\bU_1'\bU_1=\bI_p$
when $n\geq p$), we have 
$\bK'\bK=\bV\bD^2_\bullet\bV'$, $(bK'\bK)^{-1} =
\bV\bD^{-2}_\bullet\bV'$. Also $\bS_h=\bV\bOmega_h\bV'$ because it is
circulant, so that $\bS_h(\bK'\bK)^{-1}\bK' =
\bV\bOmega_h\bD_\bullet^{-1}\bU_1'$  and $\bK\bS_h(\bK'\bK)^{-1}\bK' =
\bU_1\bOmega_h\bU_1'$. Then
\begin{equation*}
  \begin{split}
    {\mathcal R}_P(\hat\bmu^h,\bmu) 
    & = \Lambda^{-2}\mathbb{E}
\parallel\!\bK\bS_h(\bK'\bK)^{-1}\bK'\by -\bK\blambda\!\parallel^2 \\
& = \Lambda^{-2}\mathbb{E}\parallel\!\bU_1\bOmega_h\bU_1'(\by-\bmu)
+\bK\bS_h\blambda -\bK\blambda\!\parallel^2 \\
& =\Lambda^{-2} \mathbb{E}\parallel\!\bU_1\bOmega_h\bU_1'(\by-\bmu) -
\bU_1\bD_\bullet(\bI_p -  \bOmega_h)\bV'\blambda \!\parallel^2.\\
& =
\Lambda^{-2}\mathbb{E}[tr\{(\by-\bmu)'\bU_1\bOmega_h^2\bU_1'(\by
-\bmu)]\\
&\qquad\qquad
 +
 \Lambda^{-2}\mathbb{E}[tr\{(\by-\bmu)'\bU_1\bOmega_h^2\bD_\bullet(\bI_p-\bOmega_h)\bV'\blambda\}]
 + \Lambda^{-2}tr\{\blambda'\bV(\bI_p-\bOmega_h)^2\bD_\bullet^2\bV'\blambda\}.\\
\end{split}
\end{equation*}
Interchanging the expectation and the trace operators, the second term vanishes 
because  $\mathbb{E}\by = \bmu$. Also using the property that the trace
of the product of two conformable matrices is the trace of their
product in the reverse order (as long as they are also conformable in
the reverse order), the first term 
equals
$ \Lambda^{-2} tr[
\bV\bD_\bullet^{-1}\bOmega_h\bU_1'\bU_1\bOmega_h\bD_\bullet^{-1}\bV'\bSigma]$
and then
\begin{equation*}
{\mathcal R}_P(\hat\bmu^h,\bmu) =
\Lambda^{-2}tr\{ \bU_1\bOmega_h^2\bU_1'\bSigma +
\blambda'\bV\bD_\bullet^2(\bI_p - 
\bOmega_h)^2 \bV'\blambda\}.
\end{equation*}
Under the idealized conditions of this section, $\bSigma\equiv
\mbox{diag}(\bmu)$ is the dispersion matrix of $\by$. Now $  {\mathcal R}_e(\hat\bff^h,\bff) = \Lambda^{-2} \mathbb{E}\parallel\!\hat\blambda^h - \blambda\!\parallel^2  =  \Lambda^{-2}
\mathbb{E}\parallel\!\bV\bOmega_h\bD_\bullet^{-1}\bU_1'\by-
\blambda\!\parallel^2$ and using similar arguments as for 
${\mathcal R}_P(\hat\bmu^h,\bmu)$ yields
\begin{equation*}
  {\mathcal R}_e(\hat\bff^h,\bff)
   = \Lambda^{-2}tr\{
\bU_1\bD_\bullet^{-2}\bOmega_h^2\bU_1'\bSigma + \blambda'\bV(\bI_p -
\bOmega_h)^2 \bV'\blambda\}  =
\Lambda^{-2}tr\{\bD_\bullet^{-2}   [\bOmega_h^2\bU_1'\bSigma\bU_1 +
  \bD_\bullet^2(\bI_p - \bOmega_h)^2 \bV'\blambda\blambda'\bV
]\}.
\end{equation*}
Exploiting the diagonality of $\bD_\bullet$ and the 
nonnegative definiteness of the matrices inside the trace operator yields
that ${\mathcal   R}'_e(\hat\bff^h,\bff) \leq
tr\{\bD_\bullet^{-2}\} {\mathcal R}'_P(\hat\bmu^h,\bmu)$. 
Using similar arguments,  ${\mathcal  R}'_P(\hat\bmu^h,\bmu) \leq
tr\{\bD_\bullet^{2}\} {\mathcal  R}'_e(\hat\bff^h,\bff)$ so that  both
risks are minimized at the same $h$. From Theorem~\ref{gcv.theorem}, 
\begin{equation*}
\begin{split}
\Lambda^{-2}\mathbb{E}\zeta(h) 
 &=  \Lambda^{-2}\{tr [(\bI_p-\bOmega_h)^2\bU_1'\bSigma\bU_1]   
 + \blambda'\bV(\bI_p - 
\bOmega_h)^2\bD_\bullet^2\bV'\blambda 
 + (1+c(h))^2tr\bU_2'\bSigma\bU_2 \} \\
& = {\mathcal
  R}_P(\hat\bmu^h,\bmu) + \Lambda^{-2}[\{1+c(h)\}^2tr\bSigma - \{2c(h)
+ c^2(h)tr\bU_1'\bSigma\bU_1 -
2tr\bOmega_h\bU_1'\bSigma\bU_1\}] \\
& = {\mathcal
  R}_P(\hat\bmu^h,\bmu) + \Lambda^{-1}[\{1+c(h)\}^2\vartheta_{n,p} -
\{2c(h)
  + c^2(h)\} tr\bvarphi_{n,p} -2tr\bOmega_h\bvarphi_{n,p}],
\end{split}
\end{equation*}
where $\vartheta_{n,p} = \Lambda^{-1}tr\bSigma$ and the matrix $\bvarphi_{n,p} =
\Lambda^{-1}\bU'\bSigma\bU$ are both free of $\Lambda$. Thus, as
$\Lambda\rightarrow \infty$, 
\begin{equation*}
\frac{\mid
\mathbb{E}\zeta(h) - {\mathcal 
  R}_P(\hat\bmu^h,\bmu) \mid }{ {\mathcal  R}_P(\hat\bmu^h,\bmu)}
\rightarrow 0.
\end{equation*}
For $n >> p$,
we have $\mathbb{E}\zeta'(h) \approx 
{\mathcal   R}'_P(\hat\bmu^h,\bmu) - \frac 2\Lambda\frac
d{dh}tr\bOmega_h\bvarphi_{n,p}$ so that for large $\Lambda$, the risk
has an inflexion point close to the bandwidth optimizing
$\mathbb{E}\zeta(h)$. On the other hand, for smaller values of
$\Lambda$, the $h$ optimizing ${\mathcal R}_P(\hat\bmu^h,\bmu)$ is
large and close to the minimizer for $\mathbb{E}\zeta(h)$. (To see
this, consider the example of using a Butterworth filter for which
the $\nu$th diagonal element of $\bOmega_h$ is 
$(1+h\parallel\nu\parallel^r)^{-1}$.) This discussion provides some
theoretical understanding of GCV's good performance in selecting $h$ 
for all values of $\Lambda$ when $n>>p$ as is the case with emission
tomography or in our experiments.

A reviewer wondered about performance when $n>>p$ is not
satisfied. The supplement shows results on our large-scale simulation
study done for cases when $\bK'\bK$ is nearly ill-conditioned, and
also not as well-conditioned as in our experiments in
Section~\ref{simulation}. Interestingly the GCV-estimated BPF methods
do not do well relative to the optimal, but the GCV-estimated BPF+
methods continue to do well. This phenomenon needs more study.

\section{Discussion}
\label{discussion}
This paper developed a computationally efficient and
practical approach to selecting the filter resolution size in 2D FBP
reconstructions. Our approach hinges on 
implementing FBP through its equivalent BPF form, uses GCV and 
outperforms available adaptive methods in simulated PET
studies, irrespective of the total expected rates of
emissions. The  approach also has the ability to incorporate
a wider class of elliptically symmetric 2D reconstruction filters with
the potential for further improving  performance.
In general, FBP is more commonly used than BPF, but this is
perhaps because of its origins in X-ray computed tomography where
reconstruction can 
begin along LORs for a given projection angle even while data along
other projection angles are being acquired. However, in emission
tomography, the data need to be
completely acquired in the given time interval before reconstruction
can begin so using BPF may not be much of a slower alternative to FBP.
The easy estimation of the filter resolution
size and its good performance even at lower emissions rates (which
translates to lower signal-noise ratio for other applications)
potentially makes it  desirable to also use BPF in applications where
reconstruction  
in the form of the 1D filtering step can be begun synchronous with data
acquisition at other projection angles also, This would hold
especially if the waiting time for  data acquisition at all angles is more than
compensated by the increased reconstruction accuracy afforded by  GCV
selection of the bandwidth.  Methods speeding up backprojection~\citep{miquelesetal18} can further reduce the cost for using BPF.

There are a number of extensions that could benefit from our
development. For instance, adopting an improved windowing function for
windowed FBP has been shown~\citep{zeng15} to improve reconstruction
accuracy over FBP. It would be instructive to see the performance of 
GCV-selected bandwidths in such scenarios. Separately, the
FORE algorithms~\citep{defriseetal97} 
recast the 3D PET reconstruction problem into several 2D
reconstructions. \citet{lartizienetal03} showed that FORE
reconstructions using ordered subsets expectation maximization
(FORE+OSEM) are out-performed by the attenuation-weighted ordered
subsets expectation maximization (FORE+AWOSEM) refinement and FORE+FBP
reconstructions. It would be worth 
investigating whether FORE+FBP reconstructions can be further
improved by using BPFe+ in place of FBP, and with optimal
GCVe-estimated bandwidth. 
It would
also be worth evaluating whether BPFe+ reconstructions with
optimal GCVe-estimated bandwidths can  improve
estimates of kinetic model parameters in dynamic PET imaging where
FBP reconstructions are the norm. There is  scope for 
optimism here, given our method's good performance for both lower and
higher radiotracer uptake values. Nevertheless, this performance needs to be
evaluated and calibrated in such contexts. Finally, another set of
potential extensions could make possible the 
practical implementation of penalized reconstruction
methods~\citep{thompsonetal89} in BPF or for  regularizing reconstructions
obtained using \citep{vardietal85,green90,osullivanandosuilleabhaein13}. 
Thus, we see that while the methods developed here show promise in
improving 2D FBP/BPF reconstruction by improved estimation of the
filter resolution size using GCV, issues that merit further attention remain.

\appendix
\subsection{Proof of Theorem~\ref{gcv.theorem}}
\label{appendix1}
The development and proof of the theorem closely mirror that of
estimating the ridge regression parameter in~\citet{golubetal79}.
Define the  $(n-1)\times p$ matrix $\bK_{-j}$ to be $\bK$ with
the $j$th row $\bk_j'$ removed. So, $\bK' =
(\bK_{-j}',\bk_j)$. We will use the following equalities: 
$\bK_{-j}'\bK_{-j} = \bK'\bK-\bk_j\bk_j'$ and $\bK_{-j}\by_{-j} =
\bK\by-\bk_jy_j$ and $\hat\blambda_{-j}^h =
\bS_h(\bK_{-j}'\bK_{-j})^{-1}\bK_{-j}\by_{-j}$. 
The
LOOCV mean squared error~(CVMSE) is 
$\tau(h) = \frac1n\sum_{j=1}^n(\bk_j'\hat\blambda^h_{-j}-y_j)^2.$
Let $\hat\blambda=\bQ_{\bK}\bK'\by$ be the
(unsmoothed) LS reconstruction, with $\bQ_{\bK}\equiv(\bK'\bK)^{-1}$
in order to compress expressions. 
Using the Sherman-Morrison-Woodbury theorem, it follows that  
\begin{equation*}
\begin{split}
  \bk_j'\hat\blambda^h_{-j}-y_j & 
= \bk_j'\bS_h\left[\bQ_{\bK}+\frac{\bQ_{\bK}\bk_j\bk_j'\bQ_{\bK}}{1-\bk_j'\bQ_{\bK}\bk_j}\right ](\bK'y-\bk_jy_j)-y_j
\\ & = \bk_j'\hat\blambda^h+\frac{\bk_j'\bS_h\bQ_{\bK}\bk_j\bk_j'\hat\blambda} {1-\bk_j'\bQ_{\bK}\bk_j}-\bk_j'\bS_h\bQ_{\bK}\bk_jy_j-\frac{\bk_j'\bS_h\bQ_{\bK}\bk_j\bk_j'\bQ_{\bK}\bk_jy_j}{1-\bk_j'\bQ_{\bK}\bk_j}-y_j\\
& = (\bk_j'\hat\blambda^h-y_j) +
\frac{\Gamma_{jj,h}}{1-\Gamma_{jj}}(\bk_j'\hat\blambda - y_j)\\
\end{split}
\end{equation*}
where $\Gamma_{jj,h}  = \bk_j'\bS_h\bQ_{\bK}\bk_j$ and
$\Gamma_{jj}  = \bk_j'\bQ_{\bK}\bk_j$. Let 
$\bDelta_h$ be
the diagonal matrix with $\Gamma_{jj,h}/(1-\Gamma_{jj})$ as the
$(j,j)$th element. Then, using the above, the CVMSE
reduces to 
\begin{equation}
\label{cvpress}
\begin{split}
\tau(h)&=\frac1n\left\{\by'\left [\bI_n-\bK\bS_h\bQ_{\bK}\bK'\right ]'
 [\bI_n-\bK\bS_h\bQ_{\bK}\bK'\right ]\by 
 +2\by'\left [\bI_n-\bK\bS_h\bQ_{\bK}\bK'\right ]'
 \bDelta_h\left [\bI_n-\bK\bQ_{\bK}\bK'\right ]\by \\
 &\qquad\qquad
\left. +\by'\left [\bI_n-\bK\bQ_{\bK}\bK'\right ]'
\bDelta^2_h\left [\bI_n-\bK\bQ_{\bK}\bK'\right ]\by \right\}.
\end{split}
\end{equation}
Let $\bD_\bullet$ be the diagonal matrix of the square root of the $p$
eigenvalues of $\bK'\bK$. Let $\bK=\bU\bD\bV'$ be the
SVD of $\bK$ with $\bU$ as in the theorem statement, $\bD$ be
$\bD_\bullet$ augmented row-wise by an $(n-p)\times p$ matrix of 
zeros, and $\bV$ have columns 
containing the right singular vectors of $\bK$. Further, let $\bW$ be the
corresponding unitary matrix that diagonalizes any 1D~(see
\citet{bellman60}) or 2D~(see Theorem~\ref{2dcirc.theo}) circulant 
matrix. Under the 
rotated generalized linear model having observations 
$\tilde\by=\bW\bU'\by$, the reconstruction problem reformulates to
estimating  $\blambda$ from
$\mathbb{E}(\tilde\by)=\bW\bD\bV'\blambda\equiv\tilde\bK\blambda$, where
$\tilde\bK=\bW\bD\bV'$.

Now $\tilde\bK'\tilde\bK = \bV\bD_\bullet^2\bV'$ so that 
$\tilde\bK(\tilde\bK'\tilde\bK)^{-1}\tilde\bK'$ and 
$\tilde\bK\bS_h(\tilde\bK'\tilde\bK)^{-1}\tilde\bK'$ are both
circulant, with each having exactly $p$ positive and non-zero
eigenvalues given by the diagonal elements of $\bI_p$ and $\bOmega_h$
respectively. Consequently 
$(\bI_n-\tilde\bK(\tilde\bK'\tilde\bK)^{-1}\tilde\bK')=\bW {\mathcal
D}_{(\bzero_p,\bI_{n-p})}\bW^*$ 
and
$(\bI_n-\tilde\bK\bS_h(\tilde\bK'\tilde\bK)^{-1}\tilde\bK')=\bW
{\mathcal D}_{(\bI_p-\bOmega_h,\bI_{n-p})}\bW^*$ 
where $\bW^*$ is the complex conjugate transpose of $\bW$, $\bzero_p$
is a $p\times p$ matrix of zeros and ${\mathcal D}_{(\bA,\bB)}$ 
denotes a block-diagonal matrix with matrices $\bA$ and $\bB$ in the diagonals. Therefore, both $\tilde\bK\bS_h(\tilde\bK'\tilde\bK)^{-1}\tilde\bK'$ and
$(\bI-\tilde\bK(\tilde\bK'\tilde\bK)^{-1}\tilde\bK')$ are circulant
(the latter is also idempotent) with
constant diagonals. In the rotated framework,
$\Gamma_{jj,h} = tr(\bOmega_h)/n$ (note that $tr(\bOmega_h)$ is $p$
times any diagonal element of $\bS_h$)
while $1-\Gamma_{jj} = (n-p)/n$, and  so $\bDelta_h =
c(h)\bI_n$. In the rotated framework, we consider the three terms
in~(\ref{cvpress}) individually. 
The first term reduces to
\begin{equation*}
    \tilde\by'[\bI-\tilde\bK\bS_h(\tilde\bK'\tilde\bK)^{-1}\tilde\bK']'[\bI-\tilde\bK\bS_h(\tilde\bK'\tilde\bK)^{-1}\tilde\bK']\tilde\by 
\equiv \by'\bU {\mathcal D}^2_{(\bI_p-\bOmega_h,\bI_{n-p})}\bU'\by =
\bz_1'(\bI_p-\bOmega_h)^2\bz_1+\bz_2'\bz_2,
\end{equation*}
while the second term 
\begin{equation*}
2\tilde\by'[\bI_n-\tilde\bK\bS_h(\tilde\bK'\tilde\bK)^{-1}\tilde\bK']'
\bDelta_h[\bI_n-\tilde\bK(\tilde\bK'\tilde\bK)^{-1}\tilde\bK']\tilde\by
 \equiv
2c(h)\by'\bU{\mathcal D}_{(\bI_p-\bOmega_h,\bI_{n-p})}{\mathcal
  D}_{(\bzero_{p,p},\bI_{n-p})}\bU\by = 2c(h)\bz_2'\bz_2
\end{equation*} with $\bzero_{r,s}$ being the $r\times s$
matrix of zeroes. The third term
\begin{equation*}
  \tilde\by'[\bI_n-\tilde\bK(\tilde\bK'\tilde\bK)^{-1}\tilde\bK']'\bDelta_h^2 
    [\bI_n-\tilde\bK(\tilde\bK'\tilde\bK)^{-1}\tilde\bK']\tilde\by 
     \equiv 
c^2(h)\by'\bU{\mathcal
  D}_{(\bzero_p,\bI_{n-p})}\bU'\by=c^2(h)\bz_2'\bz_2.
\end{equation*}
Theorem~\ref{gcv.theorem} follows, after scaling all sides by $n$. $\Box$

\subsection{Proof of Theorem~\ref{symmcirctheo}}
\label{appendix2}
Let $\ell$ be the integer part of $(p+1)/2$.
Let
$\bc=\{c_0,c_1,c_2,\ldots,c_{\ell-1},c_{\ell},c_{\ell-1},\ldots,c_2,c_1\}$ 
be the first row of $\boldsymbol{\mathcal C}$ for even $p$; the middle
term $c_\ell$ is absent for odd $p$. 
Writing the $k$th of the $p$ complex roots of unity as $\exp\{i2\pi
k/p\} = \cos (2\pi k/p) + i\sin  (2\pi k/p)$, the 
$k$th eigenvalue of $\boldsymbol{\mathcal C}$
is $d_k = \sum_{j=0}^{p-1}c_j\omega_k^j
=c_0+\sum_{j=1}^{\ell-1} c_j\cos({2\pi kj}/p) 
+c_{p/2}(-1)^\ell; 0\leq k\leq p-1$, with the last term in the
summation absent for $p$ odd.
From \citet{bellman60} or directly, an eigenvector corresponding to
$d_k$ is $\bgamma(\omega_k) =
\{1,\omega_k^1,\omega_k^2,\ldots,\omega_k^{p-1}\}$. Further,
$d_k=d_{p-k}$ for $k=1,2\ldots,\ell-1$. This means that  any symmetric
circulant matrix has two (only one for $p$ odd) real eigenvalues of
algebraic multiplicity one with 
eigenvectors given, up to constant division,  by
$\bone=\{1,1,\ldots,1\}$ and (for $p$ even)
$\pm\bone=\{1,-1,1,-1,\ldots,1,-1\}$. There are at most $\ell-1$
distinct eigenvalues of algebraic multiplicity 2:
for $1\leq k\leq\ell-1$, the eigenvectors corresponding to $d_k$ are 
$\gamma(\omega_k)$ and $\gamma(\bar\omega_k)$, where $\bar\omega_k$
is the complex conjugate of $\omega_k$. Therefore, 
$\gamma(\omega_k)+\gamma(\bar\omega_k)$ 
and  $i(\gamma(\omega_k)-\gamma(\bar\omega_k))$ are also distinct (and
real) eigenvectors that correspond to $d_k$.  
Theorem~\ref{symmcirctheo} follows. $\Box$

\ifCLASSOPTIONcaptionsoff
  \newpage
\fi



%



\section*{Acknowledgments}
The author thanks three anonymous reviewers and an Associate
Editor for helpful and insightful comments on an earlier version of
this article. He is also grateful to S. Pal for typographical error-checking
of the mathematical proofs. 

\bibliographystyle{IEEEtran}
\bibliography{references}
%




\newpage
\renewcommand\thefigure{S-\arabic{figure}}\setcounter{figure}{0}
\renewcommand\thetable{S-\arabic{table}}
\renewcommand\thesection{S-\arabic{section}}
\renewcommand\thesubsection{S-\arabic{section}.\arabic{subsection}}
\renewcommand\theequation{S-\arabic{equation}}

\section*{Supplement}
\begin{figure*}[h]
\vspace{-0.3in}
  \centering
  \mbox{
    \subfloat[]{\label{2d-rmse-gcv-129}\includegraphics[width=\textwidth]{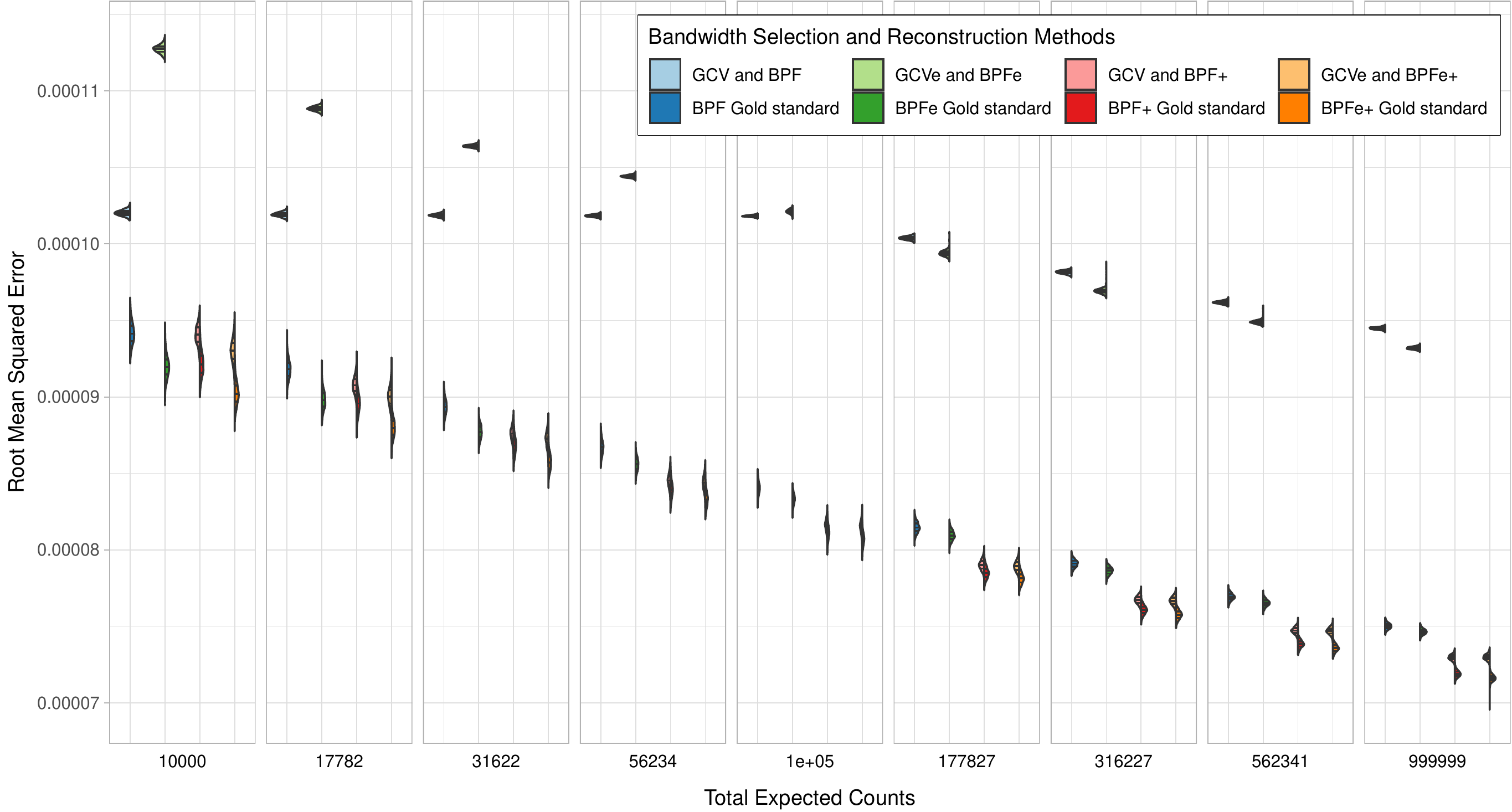}}
  }
  \mbox{
    \subfloat[]{\label{2d-rmse-gcv-160}\includegraphics[width=\textwidth]{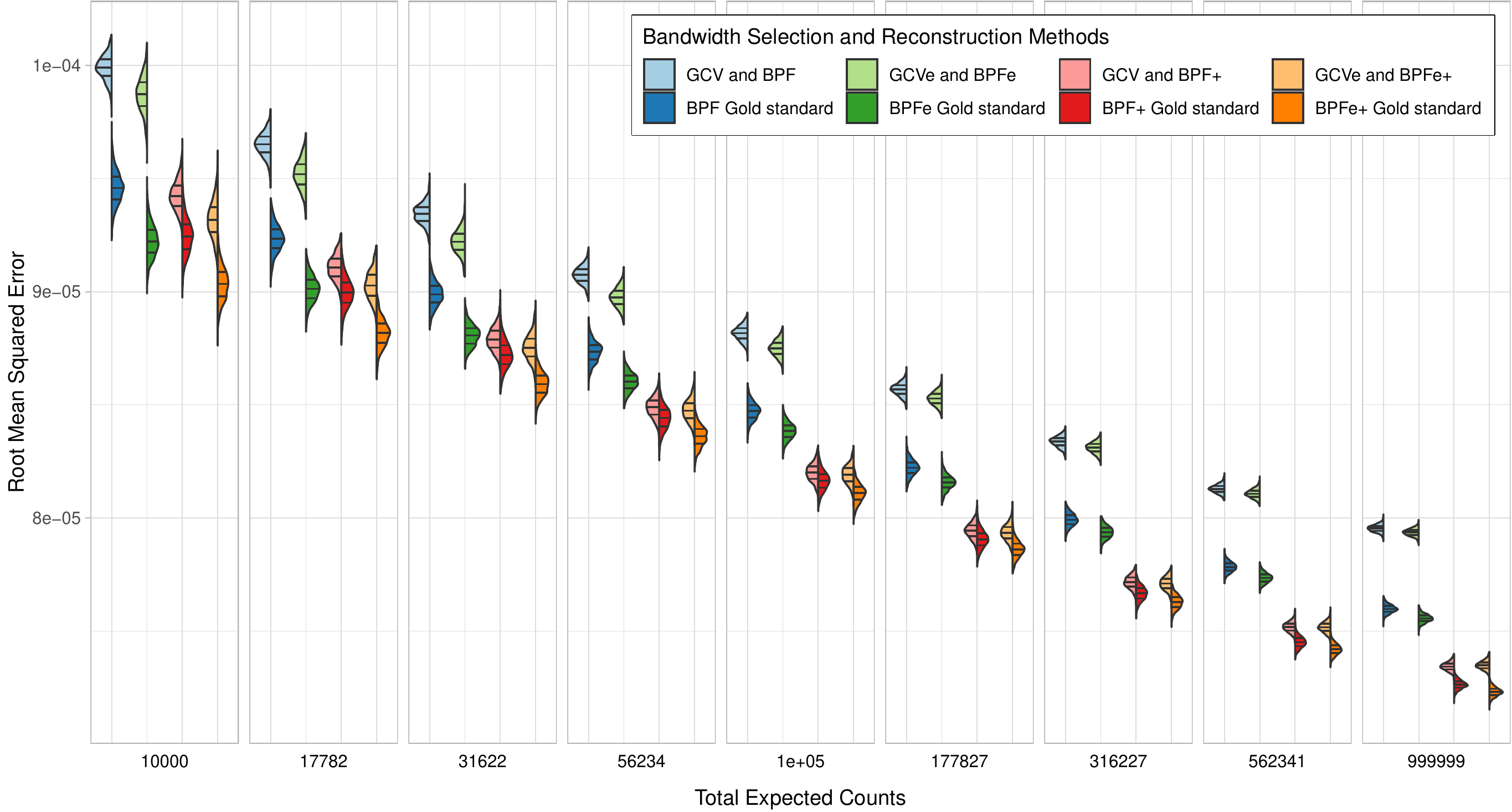}}
  }
  \vspace{-0.1in}
  \caption{Split violin plot distributions against total expected counts
  ($\Lambda$) of the 1000 RMSEs  for reconstructions using
  GCV-estimated (left violin lobe) and  the corresponding gold standard (right
  lobe) reconstructions for simulation experiments with (a) $128\times
  129$ and (b) $128\times 160$ distance-angle pairs. The imaging
  domain had $128\times 128$ pixels.
}
  \vspace{-0.1in}
\end{figure*}

\end{document}